\documentclass[reprint,twocolumn,showpacs,preprintnumbers,amsmath,amssymb,32]{revtex4}

\usepackage{multido}
\usepackage{graphicx}
\usepackage{dcolumn}
\usepackage{bm}
\usepackage{setspace}
\usepackage{xfrac}

\newcommand{\fd}{\displaystyle \frac}

\begin{document}

\title{Ultracold neutron depolarization in magnetic bottles}


\author{A.~Steyerl}
\email{asteyerl@mail.uri.edu}
\affiliation{Department of Physics, University of Rhode Island, Kingston, RI 02881, U. S. A.}

\author{C.~Kaufman}
\affiliation{Department of Physics, University of Rhode Island, Kingston, RI 02881, U. S. A.}
\author{G.~M\"uller}
\affiliation{Department of Physics, University of Rhode Island, Kingston, RI 02881, U. S. A.}
\author{S.~S.~Malik}
\affiliation{Department of Physics, University of Rhode Island, Kingston, RI 02881, U. S. A.}
\author{A.~M.~Desai}
\affiliation{Department of Physics, University of Rhode Island, Kingston, RI 02881, U. S. A.}

\begin{abstract}{We analyze the depolarization of ultracold neutrons confined in
a magnetic field configuration similar to those used in existing or proposed
magneto-gravitational storage experiments aiming at a precise measurement of the
neutron lifetime. We use an extension of the semi-classical Majorana approach as
well as an approximate quantum mechanical analysis, both pioneered by Walstrom \textit{et al}. [Nucl. Instr. Meth. Phys. Res. A 599, 82 (2009)]. In contrast with this previous
 work we do not restrict the analysis to purely vertical modes of neutron motion.
The lateral motion is shown to cause the predominant depolarization loss in a magnetic
storage trap. The system studied also allowed us to estimate the depolarization loss
suffered by ultracold neutrons totally reflected on a non-magnetic mirror immersed in
a magnetic field. This problem is of preeminent importance in polarized neutron decay
studies such as the measurement of the asymmetry parameter $A$ using ultracold neutrons,
and it may limit the efficiency of ultracold neutron polarizers based on passage
through a high magnetic field.

}


\end{abstract}

\email{asteyerl@mail.uri.edu}

\maketitle
\section{\label{sec:I} Introduction}
The neutron lifetime $\tau_{\textrm{n}}$ is an important parameter in tests
 of the Standard Model of particle physics. It also affects the rate of helium production
in the early universe and the energy production in the sun. The current
Particle Data Group (PDG) average is $\tau_{\textrm{n}}$ =  880.1$\pm$ 1.1 s \cite{1}. However,
the value of one experiment \cite{2}, which reported the lowest measurement
uncertainty of $\sim$ 0.8 s, is $\sim$ 3.5 s lower than the bulk of other data in the
PDG collection \cite{3,4,5,6,7,8}, that are grouped consistently around 882.0 s
($\pm$ 1.0 s) \cite{9}. Therefore the actual uncertainty of $\tau_{\textrm{n}}$ to be used in
cosmological calculations may be of the order of 2 seconds or more. As a
possible way of advancing this field, storage of polarized ultracold
neutrons (UCNs) in a magnetic trap has been pioneered by Paul {\it{et al.}} \cite{10}
and is currently being pursued vigorously by several groups worldwide \cite{11,12,13,14,15}.
One advantage of magnetic UCN storage versus storage in material bottles,
the method used in a number of previous neutron lifetime measurements
\cite{3,4,6,8}, is the potential absence of losses due to effects other than $\beta$-decay.
There are no wall losses, the slow loss due to quasi-stable orbits is serious
but believed to be manageable by avoiding regular orbits \cite{14}, and the potential
loss due to depolarization, defined as spin flip relative to the local field direction, is commonly assumed to be negligible. For systems
using permanent magnets the question of gradual demagnetization over time
appears to have found little attention so far.

   Until recently UCN depolarization estimates \cite{16,17} were based on Majorana's
quasi-classical result of 1932 \cite{18} for a free polarized particle with magnetic moment
moving with constant velocity vector through a non-uniform static magnetic
field of specific form. Only its spin state was assumed to be  affected by the magnetic
 field. This model predicted a depolarization probability
$ D = \exp(-\pi\omega_{L}/2\omega)$ for one passage through the field. This value decreases exponentially with the adiabaticity
parameter $\omega_{L}/\omega$,
where $\omega$ is the frequency of rotation of the field as seen from the reference
frame of the moving particle, in the critical region where the field rotates
fastest while the magnitude $B$ of the magnetic field may be small.
 $\omega_{L}$ is the Larmor frequency. For magnetic field
parameters as currently used or proposed for UCN storage, $D$ would be of
order $\exp(-10^{6})$, thus immeasurably small. Recently, Walstrom {\it{et al.}} \cite{14}
pointed out that the values of $D$ for confined, rather than freely moving,
neutrons are much larger. For a UCN moving along a vertical path in the
storage system proposed by them, $D$ was estimated to be in the range
$D \sim10^{-20}$ to $10^{-23}$. This is much larger than the Majorana value but still
negligible in any actual or projected neutron lifetime experiment.

   Using a simplified model of magnetic field distribution we extend the
theory of \cite{14} to include arbitrary UCN motion with both vertical and horizontal
velocity components, confined to the vertical space between upper and lower
turning points that depend only on the UCN energy for vertical motion. In our
model (introduced in Sec.~\ref{sec:II}) the magnetic field magnitude $B$ is uniform within any horizontal plane, so there is no horizontal component of magnetic force.
Therefore the neutron moves with constant velocity in the horizontal $z$-
and $x$- directions. We show that $D$ could reach a level approaching the tolerance
limit for a high precision neutron lifetime measurement unless precautions are
taken. As is well known the most critical issue is the choice of a
 stabilization  field perpendicular to the magnetic mirror field, of sufficient strength so that the
depolarization rate will be negligible in a neutron lifetime experiment.

   Our model field is close to the ``bathtub configuration'' of Ref.~\cite{14} but
the lateral confinement of UCNs, achieved there by double curvature of the
magnetic mirror, is simulated differently. The magnetic mirror is horizontal
and extends to infinity in both lateral dimensions. However, one could imagine the presence of ideal vertical mirrors reflecting the UCNs back and forth in
the horizontal directions without any change in the analysis.

More specifically, we use an infinite ideal planar Halbach array \cite{19}, which is free of the field ripples present in actual realizations \cite{14}. In the design of Ref.~\cite{14}, the ripples are important
only within  about 1 mm   of  the surface of the magnets.
This region is not reached by the  UCNs whose maximum energy
(for vertical motion) is $\sim$45 neV (for the parameters in \cite{14}),
 since they reverse their flight
direction before entering this zone.

   Using this model of field distribution we have also studied the
problem of depolarization of UCNs in reflection from a non-magnetic
mirror immersed in a non-uniform magnetic field. This question is
important as a mechanism that may limit the efficiency of UCN
polarizers based on transmission through a magnetic field. For a
sufficiently strong field, neutrons in only one spin state can pass the
field to proceed to the experiment. Otherwise they are reflected.
However, following the polarizer the UCNs are usually reflected on
trap or guide walls exposed to the stray field of the polarizer and
thus may lose their 100\% polarization if the reflection process involves
 depolarization. Moreover, a possible depolarization on non-magnetic
trap walls in a magnetic field is highly relevant in measurements
of the neutron decay asymmetry parameter $A$ using ultracold neutrons
\cite{20,21}. This problem has first been investigated by Pokotilovski \cite{17}
who used an adaptation of the Majorana model to the reflection geometry.
In the present work we study certain aspects of this problem by imagining
a horizontal lossless non-magnetic UCN mirror inserted at a variable
height into the magneto-gravitational storage space. The net
depolarization per bounce on this mirror will be compared to the
depolarization for one bounce in the magnetic field in the absence of
the mirror to obtain an estimate for the depolarization effect of the
mirror.

   The topic of UCN depolarization in magnetic storage or in mirror
reflection in a magnetic field raises interesting questions of quantum
interpretation. Fig.~3 of Ref.~\cite{14}
and our Fig.~\ref{fig:two} (to be discussed in Sec.~\ref{sec:III.B.2}) show the probability for the neutron to be in the
spin-flipped state (relative to the field direction) as a function of
position of the neutron as it moves through the magnetic storage space. The curve is strongly peaked at the critical
level where the field rotates fastest in the reference frame of the moving
neutron. This behavior is the same as displayed by the Majorana result
\cite{18} (where it is more difficult to deduce  since the author used a
quantization axis fixed in space rather than rotating together with the field).
In a semi-classical interpretation, as the UCN starts moving from one
turning point, say the upper one, down toward the lower one, the spin
vector rotates away from the quantization axis (which was chosen parallel
to the local magnetic field vector in Ref.~\cite{14}). It reaches a certain
maximum angle around the critical zone; then this rotation is reversed and ends at a much smaller value at the next
turning point for UCN motion. This indicates that an analyzer of neutron
polarization placed at different heights would show a variation of
depolarization by many orders of magnitude ($>$ 8 decades for the example shown in Fig.~3 of Ref.~\cite{14})
over the vertical range of the storage space. The depolarization rate
expected for an actual UCN magnetic storage experiment, without any
polarization analyzer intersecting the beam, is determined by the current
of UCNs in the ``wrong'' spin state, i.e. of  high-field seekers leaving
the system at the lower and upper turning points while
the ``correct'' (high-field repelled) state
is reflected and returns to the storage space.
 This association of net
depolarization with loss currents is consistent with the following
interpretation: At the turning points a measurement is performed
 (in the sense of quantum mechanics),
conceptually by neutron detectors placed just below the bottom and just
above the top of the storage region for a given UCN energy for vertical
motion. These detectors would intersect the UCNs in the ``wrong'' spin state as they exit the storage system. In the Copenhagen interpretation, such a measurement (actual or hypothetical) resets the UCN wave function to a pure  state of high-field repelled neutrons. The spin
state then evolves as described by the spin-dependent Schr\"odinger equation
(or its semi-classical analog) until the next ``measurement'' takes place at the
following turning point and the process of wave collapse and wave evolution is
repeated. Alternative interpretations are conceivable but we will use the
picture outlined above.

Following Ref.~\cite{14} we use the Wentzel-Kramers-Brillouin (WKB) approximation
to solve the spin-dependent Schr\"odinger equation. This appears justified since
the spatial variation of field variables (gravitational potential
and magnetic field {\bf B}) is much
slower than the variation of UCN wave function.
The scales are of order cm for gravity and {\bf B},
and of order $\mu$m or less for the neutron wavelength.

   We are aware of the fact that an exact treatment of UCN depolarization in
magnetic storage may involve quantum electrodynamics since the moving neutron,
in its reference frame, is affected by a time-dependent electromagnetic field,
i.e. by low-energy photons. We will also neglect temporal fluctuations of the
field due to mechanical vibrations or, if electromagnets are used for field
generation, AC components of the current supply. We are not aware of any work
on time-dependent effects of this kind in magnetic UCN storage.

\section{\label{sec:II} MAGNETIC FIELD DISTRIBUTION	 }

\begin{figure}[tb]
  \begin{center}
 \includegraphics[width=77mm]{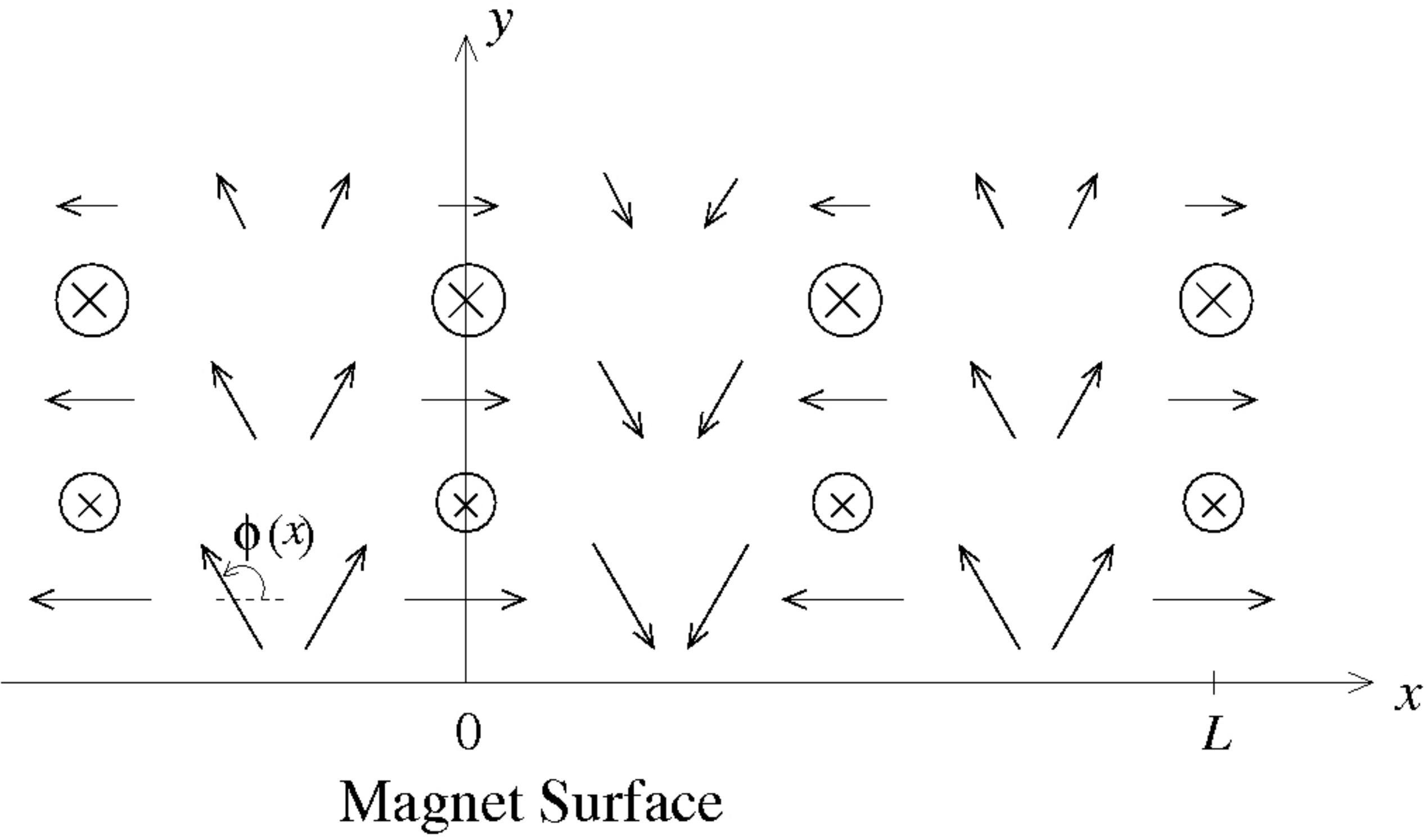}
\end{center}
\caption{For our field model, the arrows show the Halbach magnetic field
$\bm{B}_H$ as it rotates in the ($xy$)-plane. Its magnitude $B_H$ decreases
 exponentially with height $y$ and is represented by the arrow length
using a log scale. The angle $\phi=-Kx$ of the Halbach field is also shown. The superimposed stabilization field $\bm{B}_1$ in the
$z$-direction increases slowly with $y$ as in Ref.~\cite{14} and is symbolized
by the crosses of variable size.}
  \label{fig:one}
\end{figure}

We consider a Halbach array \cite{19} of permanent magnets of thickness $d$
covering the infinite $(zx)$-plane, where the $x$-axis points to the right,
the $y$-axis points up (against gravity) and
the $z$-axis toward the front (Fig.~\ref{fig:one}). We choose
 $y = 0$ at the upper magnet surface and  will closely follow the
description in Ref.~\cite{14}, apart from this choice of system of coordinates.
This choice will allow us to use the Pauli matrices in their standard form.

In the limit of infinitely fine division of magnet blocks in the $x$-direction,
let the magnetization vector have a constant magnitude $M_0$ but, viewed along
the positive $z$ direction, rotate clockwise in the $(xy)$-plane with periodicity $L=2\pi/K$ in $x$-direction:

\begin{equation}\label{(1)}
{\mathbf{M}}(x) = M_{0} ({\bm{\hat{x}}} \cos Kx + {\bm{\hat{y}}} \sin Kx).
\end{equation}

Using a complex quantity $ \bar{M} = M_x + i M_y$, Eq.~(\ref{(1)}) can also be written
$ \bar{M} = M_0 \mathrm{e}^{iKx}$. ${\bm{\hat{x}}}$ and $ {\bm{\hat{y}}}$
are unit vectors.

   We choose the same parameters as for the design in Ref.~\cite{14}, thus $L$ = 5.2 cm
and $K = 1.21$ cm$^{-1}$. The special feature of the Halbach system is that
it generates a magnetic field only on one side of the magnets, namely in the
half space $y > 0$ for clockwise sense of rotation. This Halbach field is
\begin{equation}\label{(2)}
   {\bf{B}}_{H}(x,y) = B_0  {\textrm{e}}^{-Ky} ({\bm{\hat x}} \cos Kx -{\bm{\hat y}} \sin Kx),
\end{equation}
or, in complex notation, $\bar{B}_H = B_0 {\textrm{e}}^{-Ky} {\textrm{e}}^{-iKx}$.
$B_0 = B_{\textrm{rem}} (1 - {\textrm{e}}^{-Kd})$ is determined by the remanent field
$B_{\text{rem}}$ and the block depth $d$ = 2.54 cm. The magnitude of
${\mathbf{B}}_H$, $B_H = B_0 \mathrm{e}^{-Ky}$, only depends on the vertical coordinate $y$.
 The field distribution is shown schematically in Fig.~\ref{fig:one}.

In the actual scheme \cite{14}, the uniform rotation is replaced by dividing the
rotation period $L$ into four blocks, each of length $L/4$ and with the same magnetization $M$, but with an angle of 90$^\circ$ between the directions of ${\mathbf{M}}$ in adjacent blocks (schematically represented as
 $...\leftarrow \downarrow\rightarrow\uparrow\leftarrow...)$.
Alternative designs are in the form of vertical or horizontal cylinders
where the magnets are assembled along the cylindrical surface: in a
dipolar way in \cite{11}, with adjacent blocks magnetized in the peripheral
direction with equal magnetic poles facing each other
 (schematically: $...\rightarrow\leftarrow\rightarrow\leftarrow...$). In project \cite{12} a
cylindrical octupole Halbach magnet is used where 32 blocks are distributed
uniformly over the perimeter and the direction of magnetization advances
by 56.25$^{\circ}$ from one block to the next. In these systems the magnitude $B$ of magnetic
field increases sharply near the magnetic wall. A superconducting quadrupole
system of magnetic UCN storage for a neutron lifetime experiment is used in Ref.~\cite{15} while in Ref.~\cite{13} the UCN are stored in superfluid helium using a Ioffe system with horizontal electromagnetic quadrupole. For a quadrupole the field magnitude increases
linearly with radial distance from the axis. At least one magnetic end cap
is required for all cylindrical systems; on the upper side of vertical
systems gravitational confinement can be used.

   The field distribution for systems using permanent magnets with magnetization direction
advancing in steps from block to block may be expressed as a Fourier series,
as in Eq. (7) of Ref.~\cite{14}. The first term of the expansion
is dominant and has the form (\ref{(2)}) with constant $B_0$ somewhat smaller
than $B_{\textrm{rem}}  (1 - {\textrm{e}}^{-Kd})  $. For the planar quadrupole Halbach system the reduction factor
is $4/(\pi\sqrt 2) = 0.900...$ \cite{14} and the field in the lower half
space $y < -d$  no longer vanishes. The higher Fourier
components generate a ripple field in the $(xy)$-plane, which is
significant within $\sim 1$ mm from the magnet surface (and even induces a
logarithmic divergence in the field gradient within $\sim 1\mu$m). But this
space is not accessible to the UCNs if we choose a spectrum soft
enough to ensure that all neutrons approaching the magnet from above are
reflected back up before reaching the ripple zone. We use the value
of 0.64 T for the magnetic field at a safe distance 2 mm, which
corresponds to $B_0 = 0.82$ T at the surface. Thus, for the Halbach array
generated field we assume the form (\ref{(2)}) with $B_0 = 0.82$ T and strict
confinement of the vector ${\mathbf{B}}_H$ in the $(xy)$-plane. We ignore the small field
ripple in the $z$-direction considered in \cite{14} since it also decays strongly
with distance from the magnet surface. These small perturbations are not
expected to affect the depolarization results obtained below in a significant way. Our value of $B_{0}$ is $\sim 20$\% lower than the design value of \cite{14} to take into account partial demagnetization, over time, of the NdFeB permanent magnets exposed to large fields.

   A common feature of the various magnetic UCN storage schemes is the
requirement of a bias field ${\mathbf{B}}_1$ perpendicular to the main field.
It ensures that the field magnitude $B = |{\mathbf{B}}_H + {\mathbf{B}}_1|$ exceeds a certain minimum value
everywhere in the storage volume, especially at critical positions where the
field rotates fast in the neutron's moving reference frame. The main purpose
of the present work is to provide an estimate of this minimum field for
typical field parameters, as those in Ref. \cite{14} where the
field ${\mathbf{B}}_1$ also serves the purpose of guiding the decay electrons out of
the storage space to a detector as a way to monitor the neutron decay
rate in real time. ${\mathbf{B}}_1$  is generated by a toroidal electromagnet, and it is
oriented along the longitudinal direction of the ``bathtub surface'' which corresponds to
the $z$-direction in our model with a planar, rather than curved Halbach magnet.
We use the same $y$-dependence as in Ref.~\cite{14},
${\mathbf{B}}_1 = \hat{\mathbf{z}} B_{10} \rho/(\rho - y)$
with $\rho = 1.5$ m. The magnitude of ${\mathbf{B}}_1, B_1 = B_{10} \rho/(\rho - y)$
is uniform on a horizontal plane, and since the Halbach field magnitude $B_H$ is uniform for given height $y$ no
horizontal force acts on the stored UCN.

 In Ref.~\cite{14} a value of 0.05  to 0.1 T was proposed for ${{B}}_{10}$. We will
consider field strengths down to the mT range since this range seems to be
closer to the bias field used in Ref.~\cite{11}. In this latter work the value
used was not given but it was estimated, on the basis of the Majorana
formula, that a minimum field $B_1$ of $\sim$0.001 T was required for their
neutron lifetime measurement.

   Our analysis of the evolution of spin-flip probability for UCNs moving
in our model magnetic field distribution differs from that of Ref.~\cite{14}
as follows. It is not restricted to purely vertical motion but assumes that
the UCNs can also have arbitrary horizontal velocity components $v_x$ and $v_z$.
$v_x$ and $v_z$ are constant since neither gravity nor the net magnetic
field ${\mathbf{B}} = {\mathbf{B}}_H + {\mathbf{B}}_1$ of our model exert
a horizontal force on the neutron. As noted earlier, both $B_H$ and $B_1$ are uniform at given height $y$,
and ${\mathbf{B}}_H$ is perpendicular to ${\mathbf{B}}_1$, thus
 $B = \sqrt{B_H^2 + B_1^2}$ depends
on $y$ only. As a result, the magnetic force, which is determined by
the gradient of $\mathbf{B}$, has no horizontal components and the equation
of motion is separable in three dimensions.

   We use three different methods of analysis, both for purely magnetic
confinement of UCNs with arbitrary 3D  velocity components,
and for a system involving a non-magnetic mirror placed into our model
magnetic field distribution: (a)  in Sec.~\ref{sec:III}  a quantum approach  using the
WKB approximation, (b)  in Sec.~\ref{sec:IV}  a quasi-classical  approximation and (c) in Sec.~\ref{sec:V} direct numerical
integration of the equations of motion.
In the quantum approach the stationary spin-dependent Schr\"odinger
equation
is solved using the WKB approximation, extending the method of Ref.~\cite{14} to
3D motion. The semi-classical Majorana method \cite{18} which was adapted to
magnetically confined UCNs in Ref.~\cite{14} will also be extended to 3D.
Using both methods we will also analyze UCN reflection
 on a non-magnetic mirror in a magnetic field. In Sec.~\ref{sec:V} we will
show that some  analytic results can be obtained as well by direct
numerical integration of the wave equation for the spin-flipped component
of the wave function. This is only feasible because the magnetically trapped UCNs
have relatively long wavelengths,
in the $\mu$m range, so the number of wave oscillations for the entire
integration path is not too large.

 \section{\label{sec:III}   QUANTUM MECHANICAL APPROACH   }

\subsection{\label{sec:III.A}    Basic equations}	

The wave function for a UCN moving in the magneto-gravitational field of
the trap is a linear superposition of the two eigenstates of the magnetic
moment interaction Hamiltonian

\begin{equation}\label{(3)}
\mathcal{H}_m=-\mu_{\textrm{n}} {\bm{\sigma}}\cdot{\mathbf{B},}	
\end{equation}
where $\mu_{\textrm{n}}= -1.913  \mu_{\textrm{N}}$ is the neutron
magnetic moment in terms of the
nuclear magneton $ \mu_{\textrm{N}}= 0.505\times10^{-26}$ J/T,
 ${\bm{\sigma}}$ is the Pauli spin operator, and ${\bm{B}}$ is the local magnetic field. The two eigenstates
$\chi^+$ and $\chi^-$ of $\mathcal{H}_m$ satisfy the eigenvalue equations

\begin{equation}\label{(4)}
\mathcal{H}_m \chi^{\pm}=\pm |\mu_{\textrm{n}}|B\chi^{\pm}
\end{equation}
and correspond, respectively, to neutron spin parallel to ${\mathbf{B}}$ with
spin energy $+|\mu_{\text{n}}|B$, and to antiparallel spin with energy
$-|\mu_{\text{n}}|B$ .
These spin eigenfunctions are obtained by spin rotation from the
$z$-axis to the direction of ${\bm{B}}$  through angles $\theta$  and $\phi$. The polar
field angle is  $\theta= \cos^{-1}(B_z/B) = \sin^{-1}(B_{xy}/B)$, where
$B_z = B_1$ is due to the bias field ${\bm{B}}_1$ and $B_{xy} = B_H$ is the
magnitude of the Halbach field ${\mathbf{B}}_H$. The azimuthal angle
in the $(xy)$-plane is
$\phi= \sin^{-1}(B_y/B_{xy}) = \tan^{-1}(B_y/B_x)$.

   Eq.~(\ref{(2)}) shows that for the Halbach field configuration

\begin{equation}\label{(5)}
 \phi = -Kx.
\end{equation}
Thus  $\phi$ depends only on $x$ (not on $y$ or $z$), while $\theta$ depends only on $y$.
These properties will simplify the analysis considerably. Exact
correspondence between the system of coordinates  $x,y,z$ used here
and the system $\eta,\zeta,\xi$ used in Ref.~\cite{14} (with $\zeta$ pointing up) is established if we add
the constant $\pi/2$ to the right-hand side of Eq.~(\ref{(5)}).

 Performing the spin rotation through angles $\theta$ and $\phi$ we obtain for the
spin basis vectors \cite{22} with quantization axis along ${\mathbf{B}}$

\begin{equation}\label{(6)}
\chi^- = \left( \begin{array}{c}
e_-s  \\
-c  \\
 \end{array} \right),
  \quad     \chi^+ = \left( \begin{array}{c}
c  \\
e_+s \\
 \end{array} \right),
\end{equation}
where $s = \sin(\theta/2)$, $c = \cos(\theta/2)$ and $e_{\pm} = \exp(\pm i\phi) = \exp(\mp iKx)$. We write the dependence of the wave function on position and spin in the form

\begin{equation}\label{(7)}
 \chi= \alpha^{(3)}(x,y,z) \chi^+ +  \beta^{(3)}(x,y,z) \chi^-,
\end{equation}
where we have used the superscript (3) to indicate that $\alpha^{(3)}(x,y,z), \beta^{(3)}(x,y,z)$ are functions of the three space coordinates while the corresponding functions $\alpha(y)$ and $\beta(y)$, introduced below, depend on $y$ only. $\chi$ satisfies the eigenvalue equation
\begin{equation}\label{(8)}
	E\chi =\left[ -\frac{\hbar^2}{2m}\nabla^2 + mgy + |\mu_{\textrm{n}}| {\bm{\sigma}}\cdot{\bm{B}}\right] \chi
\end{equation}
for a neutron of mass $m$ with constant total energy $E$ moving in a uniform gravitational field of magnitude $g$ and a non-uniform magnetic field $\bm{B}$. Using subscripts to denote partial differentiation, the Laplace operator acting on the wave function gives
\begin{align}\label{(9)}
&\nabla^2\chi = (\alpha^{(3)}_{xx} \chi^+ + 2 \alpha^{(3)}_x \chi^+_x + \alpha^{(3)}\chi^+_{xx} +\beta^{(3)}_{xx} \chi^- + 2 \beta^{(3)}_x \chi^-_x\nonumber \\
& + \beta^{(3)}\chi^-_{xx} )+ (x \rightarrow y) + (y \rightarrow z),
\end{align}
where for the second and third term the indicated permutations are performed. The basis vectors for quantization along the fixed $z$-axis can be expressed in
terms of the basis vectors $\chi^+$ and $\chi^-$:
\begin{equation}\label{(10)}
	           \left( \begin{array}{c}
1  \\
0 \\
 \end{array} \right)        = s e_+\chi^- + c\chi^+,   \quad     \left( \begin{array}{c}
0  \\
1  \\
 \end{array} \right)           = -c\chi^- +  s e_-\chi^+.
\end{equation}
Using Eq.~(\ref{(10)}) and noting that, from (\ref{(5)}), $\phi_x = -K$, $\phi_{xx} = 0$, $\phi_y = \phi_z = 0$
and also $\theta_x=\theta_z=0$, we obtain
\begin{align}\label{(11)}
&\chi^-_x = i s K (s \chi^- + c e_-\chi^+), \chi_x^+ = i s K (c e_+\chi^- - s \chi^+),\nonumber \\
&\chi^-_{xx} = iK \chi_{x}^-,  \chi^+_{xx} = -iK\chi_{x}^+,\nonumber \\
&\chi^-_y = \frac{1}{2}e_-\theta_y \chi^+, \chi_y^+ = -\frac{1}{2} e_+\theta_y\chi^-,\nonumber \\
&\chi^-_{yy} = \frac{1}{2} \left(\theta_{yy} e_-\chi^+ - \frac{1}{2} \theta_y^2\chi^-\right), \\
& \chi^+_{yy} = -\frac{1}{2} \left(\frac{1}{2}\theta_y^2 \chi^+ + \theta_{yy} e_+\chi^-\right),\chi^{\pm}_z = 0, \quad \chi^{\pm}_{zz} = 0. \nonumber
\end{align}
As in Ref.~\cite{14} we will use the WKB approximation \cite{23}  and keep only those terms in
Eq.~(\ref{(9)}) that contain the derivatives of the field variables ($\theta$ and $\phi$) in
lowest order since those change on the scale of centimeters
while the waves in real space, $\alpha^{(3)}$ and $\beta^{(3)}$, vary on the micrometer
scale, i.e. $\sim 10^4$ times faster. This implies that all second derivatives
of $\chi^+$ and $\chi^-$ are dropped, along with other small terms. (Using the numerical
integration described in Sec.~\ref{sec:V} we have performed test runs where the small
terms were retained. The results were the same within the precision of
numerical integration.)  Assuming that the UCN started out from a pure (+) spin state and keeping only the dominant terms in Eq.~(\ref{(9)}) we obtain
\begin{align}\label{(12)}
&\nabla^2\chi = (\alpha^{(3)}_{xx} + \alpha^{(3)}_{yy} + \alpha^{(3)}_{zz}) \chi^+ + [\beta^{(3)}_{xx}  + \beta^{(3)}_{yy} + \beta^{(3)}_{zz}\nonumber \\
&+ {\textrm{e}}^{-iKx} (-\theta_y \alpha^{(3)}_y + iK\alpha^{(3)}_x\sin\theta)] \chi^- ,
\end{align}
where we have 
used $\sin\theta = 2sc$, $\phi = -Kx$, and the fact that in practice $|\beta^{(3)}|\ll |\alpha^{(3)}|$. The functions in real space multiplying 
$\chi^+$ and $\chi^-$ can be simplified by noting that the $x$ and $z$ dependence of $\alpha^{(3)}$ has the plane wave form
$\mathrm{e}^{ik_xx}\mathrm{e}^{ik_zz}$ and $\beta^{(3)}$ is proportional to $\mathrm{e}^{-iKx}\mathrm{e}^{ik_xx}\mathrm{e}^{ik_zz}$. 
The wave numbers $k_x$ and $k_z$ are constant and $\mathrm{e}^{-iKx}$ represents a Bloch-wave modulation due to the periodicity of the 
Halbach field. In practice,  $k_x$ and $k_z$ are of order $\mu$m$^{-1}$, thus much larger than $K$ and $\theta_y$, both of which are of 
order cm$^{-1}$.

   Thus we can factor Eq.~(\ref{(12)}) in the form
\begin{align}\label{(13)}
&\nabla^2 \chi =
\mathrm{e}^{ik_xx}\mathrm{e}^{ik_zz}\{[\alpha''  -  (k_x^2+k_z^2)\alpha]\chi^+  \\
&+  \mathrm{e}^{-iKx}[\beta'' - (k_x^2+k_z^2)\beta -
(\theta'\alpha'+Kk_x\alpha\sin\theta)]\chi^-\},\nonumber
\end{align}
simplifying the notation. In Eq.~(\ref{(13)}) and henceforth, $\alpha(y)$ and $\beta(y)$
stand for the $y$-dependent parts of the wave function only,
and differentiation with respect to $y$ is denoted by primes. We also drop
the subscript $y$ from the $y$-component of the wave vector.
Thus, $\alpha^{(3)}(x,y,z) = \alpha(y)\mathrm{e}^{ik_xx}\mathrm{e}^{ik_zz}$ and
$\beta^{(3)}(x,y,z)
= \beta(y)\mathrm{e}^{-iKx}\mathrm{e}^{ik_xx}\mathrm{e}^{ik_zz}$. The terminology $\alpha(y), \beta(y)$ conforms
to that used in Ref.~\cite{14} where motion in horizontal directions was not taken
into account in the depolarization calculations.

   Inserting Eq.~(\ref{(13)}) into the eigenvalue equation (\ref{(8)}) gives \cite{14} two coupled
equations, one for spinor $\chi^+$ (i.e., for low-field seeking UCNs with spin
parallel to $\bm B$, which can be stored) and the other for $\chi^-$ (i.e., for the fraction
 of UCNs whose spin has flipped relative to $\bm B$ and which therefore can escape from the trap; the probability of flipping twice is negligible):
 \begin{equation}\label{(14)}
   E\alpha = -\frac{\hbar^2}{2m}\left[\alpha'' - (k_x^2+k_z^2)\alpha\right]+mgy\alpha+|\mu_{\textrm{n}}|B\alpha
\end{equation}
and
\begin{align}\label{(15)}
&E\beta=-\frac{\hbar^2}{2m}\left[\beta'' - (k_x^2+k_z^2)\beta-(\theta'\alpha'+Kk_x\alpha\sin\theta)\right] \nonumber \\
&+mgy\beta-|\mu_n|B\beta.
\end{align}
In the framework of the WKB approximation, the solution of (\ref{(14)}) is \cite{14}
\begin{equation}\label{(16)}
\alpha(y) = k_{+}^{-1/2}(y)\exp\Big(\pm i\Phi_+(y)\Big),
\end{equation}
where
\begin{align}\label{(17)}
&\frac{\hbar^2k_{\pm}^2(y)}{2m}= \\
&E-\frac{\hbar^2}{2m}\left(k_x^2+k_z^2\right)+mg(y_0-y)\mp|\mu_{\textrm{n}}| B(y). \nonumber
\end{align}
$k_+(y)$  is the magnitude of the $y$-component of local wave vector for
the storable (+) spin state (parallel to ${\mathbf{B}}$) and $k_-(y)$  is  that
for the ($-$) spin state.

  In Eq.~(\ref{(17)}), $y_0$ is the greatest height a neutron of energy $E$ and
given $k_x$ and $k_z$ would achieve in the gravitational field if the magnetic
field were switched off. In Eq.~(\ref{(16)}),
 \begin{equation}\label{(18)}
\Phi_+(y) =    \int_{y_s}^y k_+(u) du
\end{equation}
is the phase angle, for the + spin state,  accumulated between the start of vertical
motion  and the position $y$. The initial height $y_s$ for motion upward is
assumed to be that of the lower turning point, thus $y_{s^+} = y_l$,
and for motion downward the initial level is taken at the upper turning
point, $y_{s^-} = y_u$. The additional + or $-$ sign in the argument of the
exponential function in (\ref{(16)}), in front of $\Phi_+$, refers to this direction
of the motion; plus for upward and minus for downward, as in \cite{14}.

  The WKB wave function (\ref{(16)}) is normalized to a constant particle flux $\hbar/m$ in
the $y$-direction. For the spin-flipped UCNs, the flux in the $y$-direction is the
measure of the probability of depolarization, as shown below. At the classical
turning points, where $k_+ = 0$, the WKB form (\ref{(16)}) diverges and has to be
replaced by the Airy function, as shown in \cite{14}, but the WKB form is still
valid almost all the way to the turning point, except for the
last $\mu$m or so, since it correctly represents the asymptotic
behavior of the Airy function in this region.

  This is an important feature of the approximation used in Ref.~\cite{14} and in
the present work. It is made more explicit as follows: The asymptotic
form of the Airy function in the region of real waves (rather than the
exponentially decaying wave on the other side) is \cite{24,14}
\begin{align}\label{(19)}
&\mathrm{Ai}(-a_{s}|y-y_{s}|)\sim \left(\frac{a_{s}}{\pi k_{s}}\right)^{1/2} \cos\left(k_s|y - y_s|-\frac{i\pi}{4}\right) \nonumber \\
&= \frac{1}{2} \left(\frac{a_{s}}{\pi k_{s}}\right)^{1/2} \Big[\exp\left(ik_s|y - y_s| - \frac{i\pi}{4}\right) \\
& + \exp\left(-ik_s|y - y_s| + \frac{i\pi}{4}\right)\Big]. \nonumber
\end{align}
The wave number $k_s=\frac{m}{\hbar}(2g_{+s})^{1/2}|y-y_s|^{1/2}$ is determined by the local acceleration
$g_{+s}=\left|g+\left(\frac{|\mu_{\textrm{n}}|}{m}\right)\left(\frac{dB}{dy}\right)_s\right|$ at the turning point $y_s$, and $a_{s}=\left(\frac{m}{\hbar}\right)^{2/3}(2g_{+s})^{1/3}$. For upward motion
the WKB wave approximation (\ref{(16)}) for $\alpha(y)$, which is valid between the turning
points, is matched to the first term inside the braces of (\ref{(19)}), and for
downward motion to the second term by adjusting the constant
multiplying the Airy function.

   It follows from Eq.~(\ref{(15)}) that the wave function $\beta(x,y)$ for the spin flipped
component is determined by the inhomogeneous second-order differential equation
\begin{equation}\label{(20)}
\beta''(y) + k_-^2(y)\beta(y) = \theta'(y)\alpha'(y) + Kk_x \alpha(y) \sin\theta(y).
\end{equation}
Having separated off the $x$ and $z$ dependence allows us to choose the same WKB form for $\beta(y)$ as in Ref.~\cite{14}:
\begin{equation}\label{(21)}
\beta(y) = k_{-}^{-1/2}(y)\exp\Big(\pm i\Phi_-(y)\Big) f(y),
\end{equation}
where the function $f(y)$ modulating the WKB wave represents the amplitude
of spin flip. Apart from the modulation $f(y)$, $\beta(y)$ is
constructed in the same way as $\alpha(y)$. The phase accumulated since the start at a turning point,

\begin{equation}\label{(22)}
\Phi_-(y) =\int_{y_s}^y  k_-(u) du,
\end{equation}
always has a larger magnitude than the phase $\Phi_+(y)$ for $\alpha(y)$
since $k_-$ is greater than $k_+$ (except in zero magnetic field).

   Summarizing, the governing equation for $\beta(y)$ is the second-order differential equation
\begin{align}\label{(23)}
&\beta''(y) + k_-^2(y) \beta(y) = \theta'(y)\alpha'(y) + Kk_x \alpha(y) \sin\theta(y)\nonumber \\
& = [\pm ik_{+} \theta'(y) + Kk_x \sin\theta(y)] \alpha(y),
\end{align}
where the right-hand side represents an inhomogeneous term and $\alpha(y)$ is given by Eq.~(\ref{(16)}) while $\beta(y)$ has the form
\begin{equation}\label{(24)}
 	\beta(y) = k_{-}^{-1/2}(y)\exp\Big(\pm i\Phi_-(y)\Big) f(y).
\end{equation}

   In the second expression on the right-hand side of (\ref{(23)}) we have
carried out the differentiation of $\alpha(y)$, using the WKB rule of
considering the slow-varying terms as constant, with the result
 $\alpha' = \pm ik_+\alpha(y)$ where the + sign
applies to upward motion and the $-$ sign to downward motion. This replacement is valid except within a few $\mu$m of the turning points.

   Our Eq.~(\ref{(23)}) is the same as Eq. (28) of Ref.~\cite{14} except for the
additional, $k_x$ dependent term on the right-hand side. It is present because we include motion with finite lateral momentum $\hbar k_x$, whereas the
analysis in Ref.~\cite{14} was restricted to the special case $k_x = 0$. We will show
that this new term makes the major contribution to UCN depolarization
in magnetic storage. We also note that Eq.~(\ref{(23)}) does not depend on $k_z$,
thus motion exactly along the $z$-direction does not induce depolarization.
This is understandable since neutrons moving along the $z$-axis during
the short periods of horizontal motion at the turning points move in a
uniform ${\mathbf{B}}$-field. On the other hand, in motion along the $x$-direction
they are exposed to the strong field ripple due to the rotating Halbach
field. These features are expected to hold, on a qualitative basis, also for
the ``bathtub system'' of Ref.~\cite{14}, where the direction perpendicular
(parallel) to the curved Halbach array corresponds to our $z$-axis ($x$-axis).

Next we will solve Eq.~(\ref{(23)}) to obtain the depolarization
rate for UCN storage in our model field distribution.

\subsection{\label{sec:III.B}  Depolarization in magnetic storage  }	

\subsubsection{\label{sec:III.B.1}    Mathematical approach }  

   We will first consider a neutron of given energy and fixed values of $k_x$ and $k_z$, moving downward from the upper turning
point $y_u$. Using the WKB rule, the second derivative $\beta''$ is obtained from (\ref{(24)}) as
\begin{equation}\label{(25)}
	\beta''(y) = k_{-}^{-1/2}(y)\exp\Big(-i\Phi_-(y)\Big) \left(f'' - 2ik_- f'- k_-^2 f\right).
\end{equation}
Inserting $\beta'' (y)$ and $\alpha(y)$ from (\ref{(16)}) into Eq.~(\ref{(23)}) gives
\begin{equation}\label{(26)}
		 f''-2ik_- f' = [iU(y) + V(y)] \exp\Big(-i\Phi(y)\Big),
\end{equation}
where $\Phi =\Phi_+ - \Phi_-, U(y) = -(k_+k_-)^{1/2} \theta'$ and
$V(y) = \left(\frac{k_-}{k_+}\right)^{1/2} Kk_x \sin\theta$.

   Defining a function $F(y)$ through $f' = F \exp(2i\Phi_-)$ we obtain from (\ref{(26)})
\begin{equation}\label{(27)}
			F'(y) = [iU(y) + V(y)] \exp\Big(-i\Psi(y)\Big),
\end{equation}
where $\Psi=\Phi_++\Phi_-$. The phases $\Phi_+, \Phi_-, \Phi$ and $\Psi$ are fast-varying
quantities, while the field variables $U$ and $V$ vary slowly with $y$
and will therefore be considered constant in all differentiations.
In integrations, as needed to obtain $F(y)$ from Eq.~(\ref{(27)}), this ``WKB rule''
directly corresponds to performing the integral of products of
slow- and fast varying terms by parts and neglecting the second
term which contains the derivative of the slow-varying factor. It has
been shown numerically in Ref.~\cite{14} that for the parameters of
magnetic UCN storage at hand this procedure gives approximations
with precision in the range $10^{-4}$. Thus we obtain from (\ref{(27)})
\begin{align}\label{(28)}
&F(y)=\int_{y_s}^y[iU(y')+V(y')]\exp(-i\Psi')dy' \nonumber \\
& =\int_{\Psi_s}^\Psi \frac{iU(y')+V(y')}{k_-+ k_+} \exp(-i\Psi') d\Psi' \\
&=  \frac{i[iU(y) + V(y)]}{k_-(y) + k_+(y)} \exp\Big(-i\Psi(y)\Big). \nonumber
\end{align}
The third step in (\ref{(28)}) is an integration by parts, where only the leading term is kept.

  The lower limit of the $y$-integration in (\ref{(28)}) is the upper turning point.
Carrying out the integration in the last step of Eq.~(\ref{(28)}) we
should expect a contribution from this lower limit of integration
($y_s \approx y_u$ or $\Psi_s \approx 0$). However, such a term does not appear in (\ref{(28)})
for the following reasons:  First we note that in the quantum treatment
involving the Airy function the turning ``point'' is blurred within a range
of order $\mu$m. Second, as mentioned
in Sec.~\ref{sec:I}, we assume, as the authors of Ref.~\cite{14} did, that at a
turning ``point'' (here the region around $y_u$) the neutron starts out in a
pure low-field seeking spin state (+), i.e. from $\beta = 0$, $f = 0$. This implies that
$\alpha(y)$ and the functions $ \beta(y), f(y)$ and $F(y)$, derived from $\alpha$ and $\alpha'$ through Eq.~(\ref{(23)}), tend to 0 as $y \rightarrow +\infty$ (in
practice, for $y$ just a few $\mu$m above the classical turning point). It was
also mentioned earlier, that the WKB function used here
for $\alpha(y)$ is just the asymptotic representation
of the Airy function $\mathrm{Ai}$ which does satisfy the initial condition without any constant added since $\mathrm{Ai}$ and $\mathrm{Ai'}$ vanish for $y \rightarrow +\infty$. As a result, there is no lower-limit contribution to $F$
in (\ref{(28)}), and the same is true also for the functions
$f(y)$ and $\beta(y)$ derived below  by further integration (in equations (\ref{(29)}) and (\ref{(32)})). We can also argue
that, due to the factor $k_{+}^{-1/2}(y)$ in $V(y)$ (defined following (\ref{(26)})),
a wave containing a term derived from a finite integration constant in (\ref{(28)}) would diverge at the
endpoint $y_l$ of integration, where $k_{+}=0$, and therefore must be zero,
in the same way as in total reflection a wave increasing exponentially
inside the medium must have amplitude zero.
The singularity is avoided only by setting the integration constant in (\ref{(28)}) equal to zero.

  Remembering the definition $F = f'\exp(-2i\Phi_-)$,
we integrate Eq.~(\ref{(28)}) once more to obtain
\begin{align}\label{(29)}
&f(y) =\int_{y_s}^y F(y')\exp(2i\Phi_-') dy' \nonumber \\
&=i\int_{y_s}^y \frac{iU(y') + V(y')}{k_-+ k_+} \exp(-i\Phi') dy'= \\
&-i\int_{\Phi_s}^\Phi  \frac{iU(y') + V(y')}{k_-^2 - k_+^2} \exp(-i\Phi') d\Phi' = P(y)\exp(-i\Phi). \nonumber
\end{align}
In (\ref{(29)}), we have defined
\begin{equation}\label{(30)}
  P(y) = \frac{iU(y) + V(y)}{W(y)}, \textrm{ with } W(y) = k_-^2(y) - k_+^2(y).
\end{equation}
It follows from the definition of $k_+$ and $k_-$ in Eq.~(\ref{(17)}) that
\begin{equation}\label{(31)}
			W = k_-^2 - k_+^2 = \frac{4m}{\hbar^2} |\mu_{\textrm{n}}| B(y)
\end{equation}
depends only on the magnitude $B(y)$ of the local magnetic field.

   From the symmetry of the problem it follows that motion in the
opposite direction, from the lower turning point at $y_l$ upward to $y_u$
gives the same function $f(y)$ as in the last form of Eq.~(\ref{(29)}),
except that the phase term $\exp(-i\Phi)$ is replaced
by $\xi \exp(+i\Phi)$ and $P(y)$ is replaced by its complex
conjugate $P^*(y)$. $\xi$ is a phase factor of unit amplitude, which arises
due to the shift of reference point for $\Phi$ from $y_u$ to $y_l$
when we change from downward to upward motion. None of these differences
affect the squared magnitude $|f(y)|^2$ which measures the
probability to find the neutron in the spin-flipped state at height $y$. An explicit expression for $\xi$ will be given following Eq.~(\ref{(44)}).

   We can compare our Eq.~(\ref{(29)}) for the depolarization amplitude $f $
with the corresponding result in equations (31-35)
of Ref.~\cite{14}, where only the case $k_x = 0$ was analyzed. This corresponds
to setting $V(y) = 0$ in our analysis. Apart from this difference, our
Eq.~(\ref{(29)}) can be obtained from Eq. (35) of \cite{14} by multiplying the latter
by the factor $2ik_+/(k_-+ k_+)$. The magnitude of this factor
is close to unity if $k_+(y)$ is only slightly less than $k_-(y)$.
For fairly high-energy UCNs this is the case for most of the
path between the turning points, but not near these points. This minor difference appears to be due to the neglect, in Ref.~\cite{14}, of $f''$ in the derivation of their Eq. (31).

  The main difference between the results of \cite{14} and our numbers,
obtained below, is due to the restriction of the previous work to
$k_x = 0$. Our analysis of the depolarization current and depolarization
rate for magnetically stored UCNs yields a loss $\sim$10 decades larger
for a typical velocity $v_x$ up to $\pm3$ m/s than
the range of values, $10^{-20}$ to $10^{-23}$ given in Ref.~\cite{14} for $v_x = 0$.

\subsubsection{\label{sec:III.B.2}   Interpretation in terms of loss current and depolarization rate} 


   We can now insert $f(y)$ from (\ref{(29)}) into (\ref{(24)}) to determine the wave
function $\beta(y)$ for spin-flipped UCNs. For the downward motion, this gives
\begin{equation}\label{(32)}
		\beta(y) = k_{-}^{-1/2}(y)P(y) \exp\Big(-i\Phi_+(y)\Big).
\end{equation}

The phase $\Phi_+$ (with the index $
+$) indicates that this wave for
the ($-$) spin state propagates, not with wave number $k_-$, but
with the same wave number $k_+$ as the (+) spin state, as  it should. Using $\beta''(y) = -k_+^2(y) \beta(y)$ we can verify that the function (\ref{(32)}) solves equation (\ref{(23)}), starting from a pure (+) spin state, as described following Eq.~(\ref{(28)}).

On a more formal basis, Eq.~(\ref{(32)}) represents a particular solution to (\ref{(23)}) and we could add to (\ref{(32)}) any solution $\beta_{h\pm}(y)$ of the homogeneous equation ${\beta}_{h}''(y) + k_-^2(y) \beta_h(y) = 0$ corresponding to (\ref{(23)}). In the WKB framework, these solutions are $\beta_{h\pm}(y)=C_{\pm}k_{-}^{-1/2}(y)\exp\Big(\pm i\Phi_{-}(y)\Big)$ with arbitrary constants $C_{\pm}$. These functions represent a constant current in the upward (downward) direction for the + ($-$) sign. Thus the same current enters and leaves the storage space, resulting in a zero contribution to the net flux out which corresponds to the depolarization loss as described below. As an example, such a homogeneous term may represent a neutron that has undergone a spin flip on the way up, proceeds past $y_u$ until it reaches its reversal point in the gravitational field and, on its way down, traverses the storage space without contributing to further depolarization.

Reverting to solution (\ref{(32)}) without added terms, we associate the net depolarization over the path from upper turning point
$y_u$ to $y_l$ with the current of spin-flipped UCNs at the endpoint $y_l$, which
is consistent with the interpretation in Ref.~\cite{14}. This current represents the net flux out of the storage space since no flux enters at $y_u$.

\begin{figure}[tb]
  \begin{center}
 \includegraphics[width=77mm]{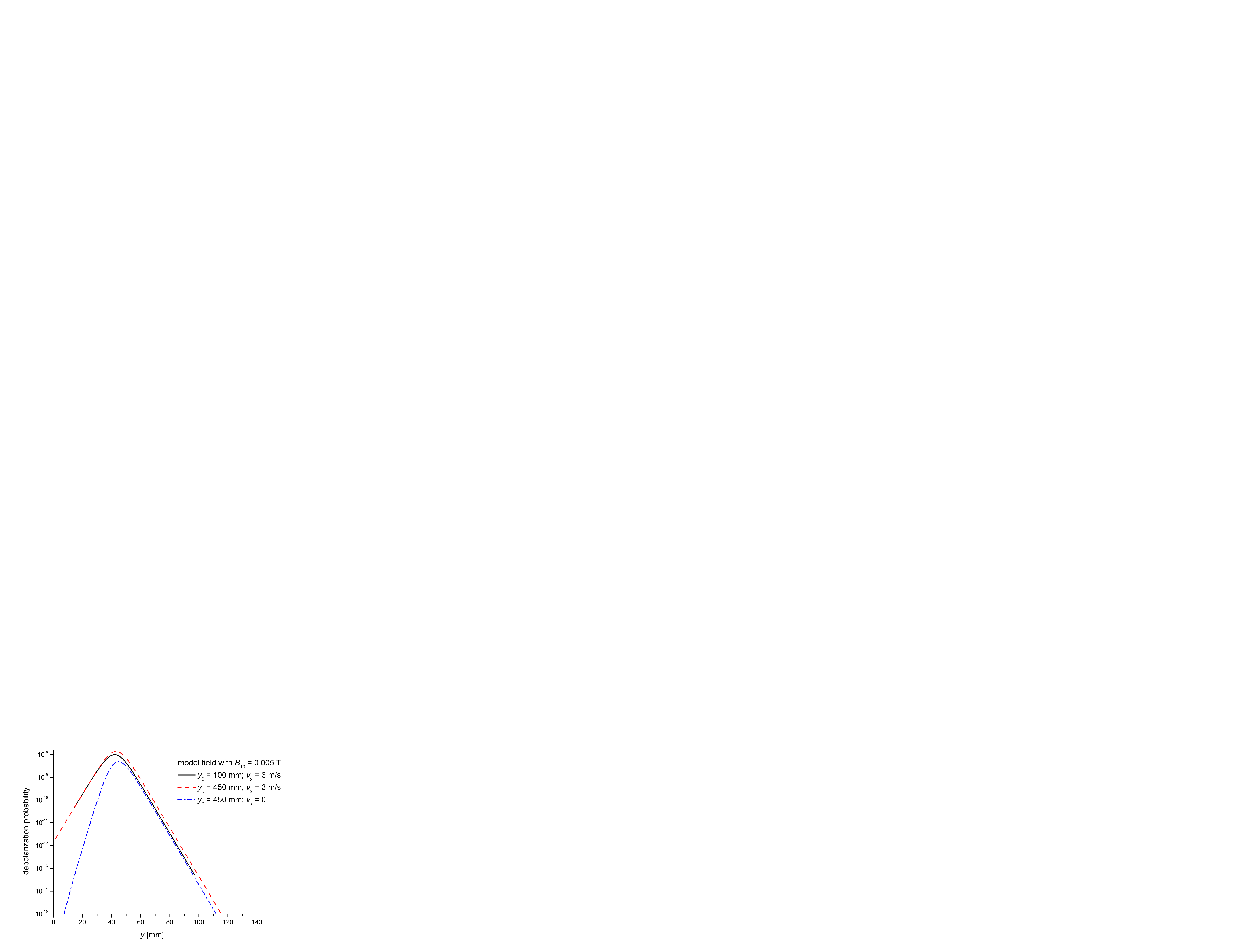}
\end{center}
\caption{(color online) Depolarization probability, given by Eq.~(\ref{(34)}) multiplied
by $m/\hbar$, as a function of neutron position for drop heights
$y_0$ = 450 mm and 100 mm, stabilization field parameter
$B_{10}$ = 0.005 T, and neutron velocity component $v_x$ = 3 m/s
or zero. The sharp peak occurs in the region where the
gradient of field angle $\theta$ is largest.}
  \label{fig:two}
\end{figure}

At an arbitrary position $y$ along the way the current $j_-(y)$ is given by \cite{22}
\begin{equation}\label{(33)}
	j_-(y) = \frac{\hbar}{m} \textrm{Re}\left[i {\beta}^*(y) \left(\frac{d\beta}{dy}\right)\right].
\end{equation}
For function (\ref{(32)}) we have
\begin{align}
&\beta^*(y) = k_{-}^{-1/2}(y)P^*(y) \exp\Big(i\Phi_+(y)\Big), \nonumber \\
&\frac{d\beta}{dy}= -ik_{+}(y) k_{-}^{-1/2}(y)P(y) \exp\Big(-i\Phi_+(y)\Big),
\nonumber
\end{align}
and thus the current as a function of position $y$ between $y_l$ and $y_u$ becomes
\begin{equation}\label{(34)}
		j_-(y) = \frac{\hbar}{m} \left(\frac{k_+}{k_-}\right) |P|^2 = \frac{\hbar}{m} \frac{k_+^2 {\theta'}^2 + K^2k_x^2\sin^2\theta}{{(k_-^2 - k_+^2})^2}
\end{equation}

   The function $(m/\hbar) j_-(y)$ corresponds to the depolarization probability of
Ref.~\cite{14}. It is plotted in Fig.~\ref{fig:two} for UCNs with energy for vertical motion determined by
the ``drop heights'' $y_0 = 10$ cm and 45 cm ($y_0$ was defined following Eq.~(\ref{(17)})), a
bias magnetic field $B_{10}$ = 0.005 T and $v_x$ = 3 m/s. As in Fig.~3 of \cite{14} we see
a sharp peak at the $y$-position where $\theta'$ is large, and a decrease as the particle
drops further down. The third curve in Fig.~\ref{fig:two} is for $y_0$ = 45 cm,
$B_{10}$ = 0.005 T and $v_x$ = 0. The peak value and the decrease on the
upper side are quite similar. Below the peak position the curve
for $v_x$ = 0 decreases faster than for $v_x$ = 3 m/s.

   The current leaving the storage space at $y = y_l$ is
\begin{equation}
j_l = \frac{\hbar}{m} \left(\frac{k_{+l}}{k_{-l}}\right) |P_l|^2 = \frac{\hbar}{m} \frac{k_{+l}^2 {{\theta}'_l}^2 + K^2k_x^2\sin^2\theta_l }{(k{_{-l}}^2 - k_{+l}^2)^2}
\nonumber
\end{equation}

\begin{equation}\label{(35)}
= \frac{\hbar}{m} \frac{K^2k_x^2}{k_{-l}^4}\sin^2\theta_l ,
\end{equation}
where the index $l$ refers to the values at $y = y_l$ and the last
form of Eq.~(\ref{(35)}) uses the fact that $k_+$ vanishes at the turning
points. In the analysis of Ref.~\cite{14} for $k_x = 0$ only this vanishing term ($\sim k_{+l}^2 {{\theta}'_l}^2$) appeared, and a higher order of approximation as well as numerical
integration were used to estimate the depolarization probability.
The result was a very small value which is negligible compared to the
second term given in the last form of (\ref{(35)}), even for values
of $v_x$ as small as  0.1 m/s.

The dependence of (\ref{(35)}) on the primary field variables is established by noting that $\sin^{2}\theta=B_{H}^{2}/B^{2}$, $k_{-l}^{4}\sim B_{l}^{2}$ and $K^{2}k_{x}^{2}=\left(\frac{m}{\hbar}\right)^{2}\omega^{2}$, where $\omega = 2\pi v_{x}/L$ is the frequency of the Halbach field as seen by the moving UCN.

   For upward motion from $y_l$ to $y_u$ we get the same result for the
current as in (\ref{(35)}) except that all indices $l$ are replaced by $u$, i.e., the quantities
relevant for the spin-flipped current leaving the system at the upper
turning point are determined by the field angle $\theta_u$ and by $k_{-u}$ at $y_u$.

   The combined depolarization loss for one reflection on the magnetic
field, i.e. for one complete round trip down and up, is determined by
\begin{equation}\label{(36)}
\frac{m}{\hbar}  (j_l + j_u) = K^2k_x^2\left(\frac{\sin^2\theta_l}{k_{-l}^4} + \frac{\sin^2\theta_u}{k_{-u}^4}\right).
\end{equation}
To approximate the actual situation in magnetic UCN storage, where the
UCNs have positive and negative velocities in any direction and, for a
low-energy Maxwell spectrum, with uniform probability per unit
of $k_x, k_y$ and $k_z$ (since the phase space density is constant),
we take the mean value of $k_x^2$ in (\ref{(36)}) for the spectral
interval $-k_{x,\textrm{max}} < k_x < +k_{x,max}$ with the result
\begin{equation}\label{(37)}
   \frac{m}{\hbar} \langle j_l + j_u\rangle = K^2\left(\frac{k_{x,\textrm{max}}^2}{3}\right) \left(\frac{\sin^2\theta_l}{k_{-l}^4} + \frac{\sin^2\theta_u}{k_{-u}^4}\right).
 \end{equation}

  As a final step in this analysis of depolarization in the WKB approximation
we establish the explicit connection between the loss current (\ref{(37)}) and the
rate of depolarization, $\tau_{\textrm{dep}}^{-1}$, that is observable as a contribution to
the decay rate (but should be negligible compared to the neutron $\beta$-decay
rate in a neutron lifetime measurement).
For given neutron energy for vertical motion, i.e. fixed turning levels
at $y_l$  and $y_u$, the depolarization rate (in s$^{-1}$) is determined by the loss
current (\ref{(37)}) divided by the number of UCNs in the field-repelled
spin state present in the trap,
\begin{equation}\label{(38)}
N = 2 \int_{y_{l}}^{y_u} |\alpha(y)|^2 dy = 2 \int_{y_l}^{y_u} \frac{1}{k_+(y)} dy.
\end{equation}
We have used the square magnitude of the WKB form (\ref{(16)})
for $\alpha(y)$ as the density. The factor 2 takes into account that both
downward and upward moving UCNs are in the trap at the same time.

 Since $k_+ = (m/\hbar) v_+$ and $dy = v_+ dt$, the expression in (\ref{(38)})
equals $(\hbar/m)T$ where $T$ is the time required for one round trip down and
up. Thus, the depolarization rate is
\begin{align}\label{(39)}
&\tau_{\textrm{dep}}^{-1} = \frac{\langle j_{l }+ j_{u}\rangle}{N}=\frac{m}{\hbar} \frac{\langle j_{l} + j_{u} \rangle}{T}\nonumber \\
& = K^2 \left(\frac{k_{x,\textrm{max}}^2}{3}\right) \left(\frac{\sin^2\theta_ {l}}{k_{-l}^4} + \frac{\sin^2\theta_{u}}{k_{-u}^4}\right)\frac{1}{T} ,
\end{align}
where we have inserted (\ref{(37)}) for the current in the last step. This shows that
the loss current (\ref{(37)}) of spin-flipped UCNs is the loss per round trip, i.e. for one
bounce in the magnetic field. This interpretation is consistent with the
interpretation in Ref.~\cite{14}.

   To compare to actual experiments storing polarized (+) UCNs in a broad velocity
range in three dimensions, we have to average (\ref{(39)}) also over $v_z$ and $v_y$.
Averaging over $v_z$ is trivial since (\ref{(39)}) does not depend on $v_z$.

   Averaging over $v_y$ can be achieved as follows. As a measure of $v_y$
for a stored UCN we could take its value at any height within the
confinement range, but the most convenient choice of reference
plane is the neutral plane at $y = y^{(\textrm{n})}$ where the gravitational force
is compensated by the magnetic force pushing upward, i.e.
where $|\mu_{\textrm{n}}| \frac{dB}{dy} = -mg$. This is the plane where the UCNs with the
lowest energy for vertical motion reside. In our field model, a UCN
with vertical velocity $v_y^{(n)} = 0$ in the neutral plane floats or moves along
the plane at constant speed. In actual confinement fields as in \cite{14} they
would follow closed or open paths on the curved neutral surface.
For small values of $v_+^{(n)}$ the vertical motion is a classical harmonic oscillation with natural frequency $\omega_0 = \left(\frac{dg_+}{dy}\right)^{1/2}$
where $g_+ = g + \frac{|\mu_{\textrm{n}}|}{m} \frac{dB}{dy}$ is the net downward acceleration.
 This implies that for small oscillations about the
neutral plane the time for a round trip becomes
$T = 2\pi/\omega_0 = 2\pi\left(\frac{dg_+}{dy}\right)^{-1/2}$. For larger
vertical velocities the oscillator potential is strongly anharmonic but
the drop height $y_0$, used originally as a measure of energy for vertical
motion, is unambiguously determined by $v_+^{(\textrm{n})}$. Therefore, if we plot
the depolarization rate (\ref{(39)}) versus $v_+^{(\textrm{n})}$, rather than $y_0$, the mean
height of this curve in the range from $v_+^{(n)} = 0$ to its maximum value
for the stored UCN spectrum directly gives the average value of
depolarization rate for a Maxwell spectrum. (The trap loading process
used in an actual experiment may induce deviations from the Maxwell spectrum.)

\begin{figure}[tb]
  \begin{center}
 \includegraphics[width=77mm]{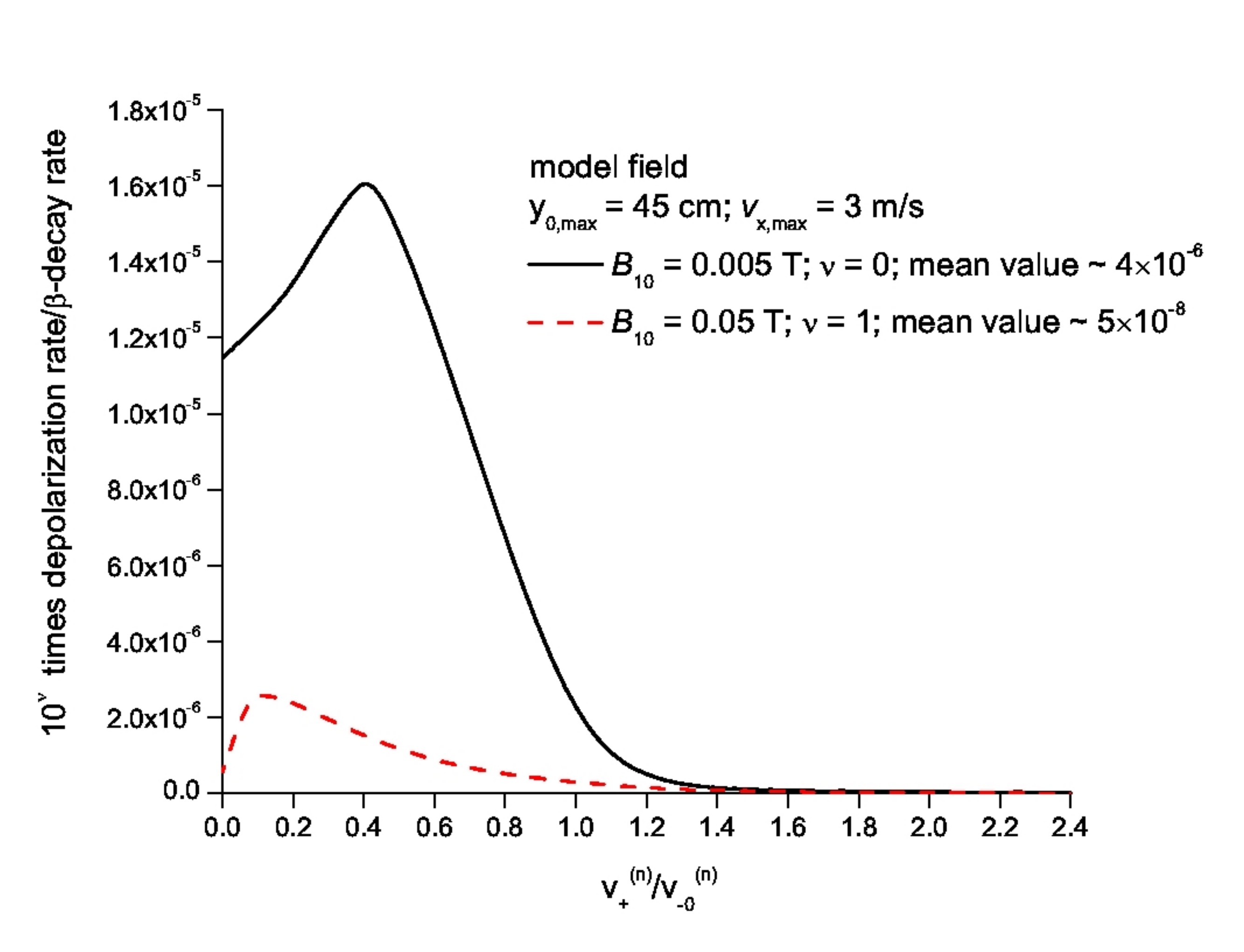}
\end{center}
\caption{(color online) Ratio between mean depolarization rate, given by
Eq.~(\ref{(39)}), and neutron $\beta$-decay rate (for a lifetime of 882s),
plotted as a function of vertical velocity component $v_+^{(\textrm n)}$ in
the neutral plane (where the gravitational and magnetic forces
are balanced). $v_+^{(\textrm n)}$ is normalized with the constant
$v_{-0}^{\textrm{(n)}}$ which
is determined by the field magnitude in the neutral plane. The
curve for $B_{10}$ = 0.005 T is plotted to scale $(\nu= 0)$ and the curve
for $B_{10}$ = 0.05 T is plotted with magnification factor $10^1$ $(\nu= 1)$.
Their difference by about two orders of magnitude shows the
strong suppression of depolarization by a stabilization field of
sufficient strength. For a Maxwell spectrum, the mean height of
the curves over the range  of the abscissa,
from 0 to 2.5 for $B_{10}=0.05$T and from 0 to 4.7 for $B_{10}=0.005$T,
 directly
determines the average over the full spectrum (here for
 $-3$ m/s $<v_x<$ $+3$ m/s and drop heights $y_0$ up to 450 mm).}
  \label{fig:three}
\end{figure}

 Such a plot is presented in Fig.~\ref{fig:three} where we have normalized
$v_+^{(\textrm{n})}$ to  $v_{-0}^{(n)}$, the  $y$-velocity for the spin-flipped state on the neutral plane for  $v_+^{(\textrm{n})}=0$.
 $v_{-0}^{(\textrm{n})} = 2 \left(\frac{|\mu_{\textrm{n}}{|B^{(\textrm{n})}}}{m}\right)^{1/2}$
 is solely
determined by the field magnitude $B^{(n)}$ on the neutral plane.
The parameters used are: $y_{0,\textrm{max}} = 45$ cm, $B_{10} = 0.005$ T and $0.05$ T and
 $v_{x,\textrm{max}} = 3$ m/s. For these parameters the mean
depolarization rate, normalized to the $\beta$-decay rate $1/\tau_{\textrm{n}}$,
is $\tau_{\textrm{n}}\langle \tau_{\textrm{dep}}^{-1}\rangle = 4\times 10^{-6}$ for
$B_{10}=0.005$ T and about two orders of magnitude less for $B_{10}=0.05$ T.

It might come as a surprise that the largest contribution to
the depolarization rate originates from UCNs with fairly low energy of vertical motion. They move through the
field almost horizontally, with small vertical oscillations about the neutral
plane. The result is plausible since these UCNs spend the largest fraction of
time in the region where the field  rotates rapidly in the reference
frame of the laterally moving neutron.

   In section \ref{sec:IV.B} we will analyze the same problem of depolarization in
magnetic storage of UCNs using the semiclassical method and compare
the two approaches. Next we apply the quantum
approximation to analyze the depolarization in reflection of polarized
neutrons on a non-magnetic mirror immersed in a non-uniform
magnetic field.

\subsection{\label{sec:III.C}  Reflection on a non-magnetic mirror in a magnetic field }	

\begin{figure}[tb]
  \begin{center}
 \includegraphics[width=77mm]{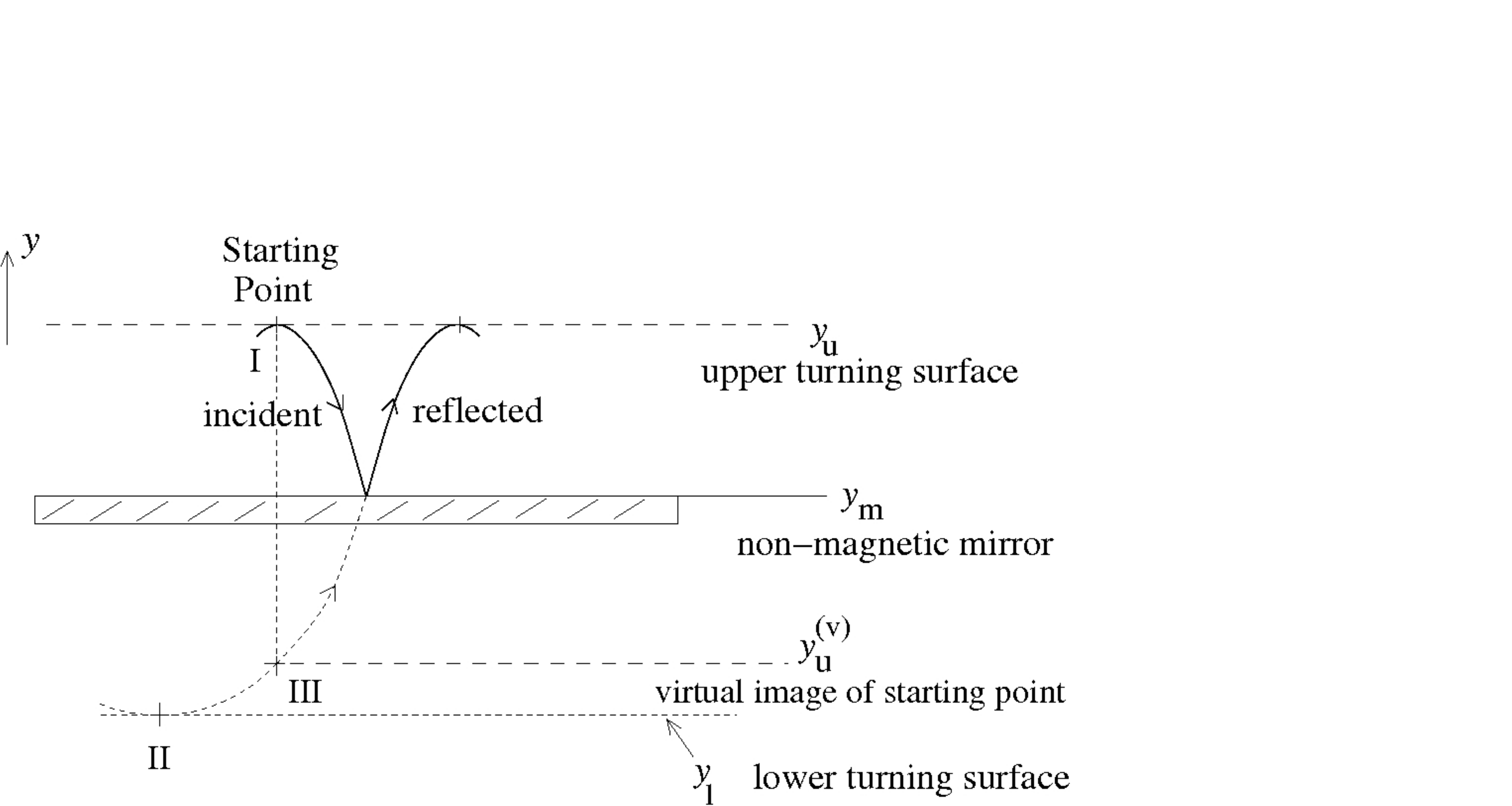}
\end{center}
\caption{Geometry of non-magnetic mirror conceptually placed
into the magnetic storage space at various levels $y_m$ between
the upper and lower turning surfaces at $y_u$ and $y_l$ for a given
UCN energy for vertical motion. The reflected spin-flipped wave
consists of the two components in Eq.~(\ref{(42)}): (a) the particular
solution $\beta_p$ which is the same as for a wave moving upward
from the lower turning point at II in the absence of the mirror;
(b) the homogeneous wave $\beta_h$ induced at the mirror surface by
the wave incident from above, starting from the upper turning
level at point I. The position III of the virtual image of I below
the mirror determines a phase angle.}
  \label{fig:four}
\end{figure}

   In Sec.~\ref{sec:I} we mentioned two examples where the possible
depolarization of UCNs in total reflection from a non-magnetic wall, like
copper, is of major importance: (a) UCN experiments on spin anisotropy
parameters in neutron decay, such as the neutron-electron spin
correlation coefficient  $A$ \cite{20,21}; and (b) UCN polarizers based on
transmission through a high magnetic field. Depolarization in mirror
reflection has been investigated theoretically in Ref.~\cite{17}. We will
study aspects of this problem by imagining an ideal, non-magnetic mirror
inserted horizontally into our magnetic model field at various heights $y_m$, as shown in Fig.~\ref{fig:four}, 
and comparing the depolarization per bounce on this mirror with that in the
field without the mirror. The reflecting mirror surface at $y_m$ lies between the
upper and lower turning points $y_u$ and $y_l$ for UCNs. We use index $m$ to denote
the quantities at the reflecting mirror surface and have assumed
that the UCNs have given lateral velocities $v_x$ and $v_z$ and a fixed energy
for vertical motion.

An ideal UCN mirror has reflection amplitude $R = \exp(-2i\Sigma)$ and reflectivity $|R|^2 = 1$. The phase angle 
$\Sigma = \cos^{-1}(k_{+m}/k_c)$ depends on the limit $k_c$ of total reflection and, for the system at hand, on the vertical component of 
incident wave vector, $k_{+m}$, which is the same for both wave components, $\alpha$ and $\beta$. We will see that the depolarization at the 
mirror is independent of $\Sigma$, i.e., it should be virtually the same for any low loss (almost ideal) mirror material. On physical grounds, 
no abrupt change of depolarization probability $|\beta|^2$ is expected since the dwell time of $\sim 10^{-8}$ s inside the wall is much 
shorter than the Larmor precession period (1 $\mu$s for $B=34$ mT) which sets the time scale for any change.

      As in the previous sections, we assume
that the particle started out from an upper turning level $y_u$ in a pure (+) spin
state. On incidence at $y = y_m$ its wave function has acquired a
depolarized component given by Eq.~(\ref{(32)}):
\begin{equation}\label{(40)}
  \beta(y_m) = k_{-}^{-1/2}(y_m)P(y_m) \exp\Big(-i\Phi_+(y_m)\Big)
\end{equation}
where $P$ has been defined in (\ref{(30)}). The evolution of $\beta(y)$ following reflection
is determined by the inhomogeneous differential equation (\ref{(23)}):
\begin{equation}\label{(41)}
\beta''(y) + k_-^2(y)\beta(y) = [ik_{+} \theta'(y) + Kk_x \sin\theta(y)] \alpha(y),
\end{equation}
where we replaced the $\pm$ sign in (\ref{(23)}) by + since the reflected wave is propagating upward.

   The solution to (\ref{(41)}) can be written as a superposition
\begin{equation}\label{(42)}
				\beta(y) = \beta_p(y) + \beta_h(y)
\end{equation}
of a particular solution, $\beta_p(y)$, and the general solution $\beta_h(y)$ of the
homogeneous equation
\begin{equation}\label{(43)}
{{\beta}_{h}''}(y) + k_-^2(y) \beta_h(y) = 0.
\end{equation}
The amplitude multiplying $\beta_{h}(y)$ is adjusted to satisfy initial conditions.

   A particular solution $\beta_p$ for downward motion has been given
as Eq.~(\ref{(32)}). For the mirror-reflected beam we need the corresponding
solution for upward motion. It follows from the discussion
in Sec.~\ref{sec:III.B.1} that this particular solution is
\begin{equation}\label{(44)}
\beta_p(y) = \xi k_{-}^{-1/2}(y) P^*(y) \exp\Big(i\Phi_+(y)\Big).
\end{equation}
The phase factor $\xi = \exp(-i\Phi_t)$, with
$\Phi_t =  \int_{y_{u}}^{y_{l}} (k_+ - k_-) dy$,
 arises as follows:
For the mirror reflection problem we choose a time axis with $t = 0$ at the
upper turning point $y_u$ and increasing as the motion proceeds. However, the
corresponding time $t'$, chosen in Sec.~\ref{sec:III.B.1} for the upward motion, starts
from $t' = 0$ at $y_l$ (not from $y_u$). The times required to reach a given particle
position are related through $t' = -t + \frac{T}{2}$, where, as before, $T$ is the time
for a round trip down and up. The phase factor $\xi$ arises due to this
difference in the time coordinates.

   The homogeneous equation (\ref{(45)}) has two solutions but only
one, $\beta_h(y)\sim \exp\Big(+ik_{-m}(y - y_m)\Big)$, corresponds to upward
wave propagation, as required for the reflected wave. The plane wave
form is valid only near the mirror surface where $k_-(y)$ is considered constant
on the scale of the neutron wavelength. Thus, in the framework of
the WKB approximation, we use the homogeneous solution
\begin{equation}\label{(45)}
 \beta_h(y) = C k_{-}^{-1/2}(y)\exp\Big(+i\Phi_-(y)\Big)
\end{equation}
where the constant $C$ is to be adjusted to match the outgoing wave (\ref{(42)}) to the wave
 \begin{equation}\label{(46)}
 \beta_m = r k_{-}^{-1/2}(y_m)P(y_m) \exp\Big(-i\Phi_+(y_m)\Big)
\end{equation}
excited by the incoming beam. Expression (\ref{(46)}) is the
incoming wave (\ref{(32)}) multiplied
by the phase factor $r = R \exp(-2i\Phi_{-m})$ where $R$ is the reflection
amplitude. The factor $\exp(-2i\Phi_{-m})$ is due to a shift of time scales similar to
that defined following Eq.~(\ref{(44)}) but now referring to mirror reflection with
start from the upper turning point $y_u$ versus start from its virtual image
 at $y_u^{(v)}$ below the mirror surface, as shown in Fig.~\ref{fig:four}.

   Now we match the outgoing wave $\beta_p(y) + \beta_h(y)$ (from (\ref{(44)}) and (\ref{(45)}))
to the wave $\beta_m$  at the mirror surface $y_m$ (from (\ref{(46)})) to determine
the constant $C$, with the result
\begin{equation}
C = (\beta_m - \beta_{pm}) \exp(-i\Phi_{-m}) =  [rP_m \exp(-i\Phi_{+m})
\nonumber
\end{equation}
\begin{equation}\label{(47)}
 - \xi P^*_m \exp(+i\Phi_{+m})] \exp(-i\Phi_{-m}).
\end{equation}
Thus the mirror reflected wave becomes
\begin{align}\label{(48)}
&\beta(y) = \beta_p(y) + \beta_h(y) = k_{-}^{-1/2}(y)\Big\{\xi P^*(y) \exp\Big(i\Phi_+(y)\Big) \nonumber \\
&+\Big[rP_m \exp(-i\Phi_{+m})\\
&-\xi P^*_m \exp(+i\Phi_{+m})\Big] \exp(-i\Phi_{-m}) \exp\Big(i\Phi_-(y)\Big)\Big\}. \nonumber
\end{align}
The first term in the braces represents an outgoing wave with wave
number $k_{+}$ and the second one with wave number $k_-$. To calculate
the outgoing current we also need the derivative
\begin{align}\label{(49)}
&\frac{d\beta}{dy} = k_{-}^{-1/2}(y) \Big\{ik_+ \xi P^*(y) \exp\Big(i\Phi_{+}(y)\Big)
\nonumber\\
&+ik_-\Big[rP_m \exp(-i\Phi_{+m})\\
&-\xi P^*_m \exp(+i\Phi_{+m})\Big]\exp(-i\Phi_{-m}) \exp\Big(i\Phi_-(y)\Big)\Big\}. \nonumber
\end{align}
The current propagating in the upward direction is obtained from \cite{22}
\begin{equation}\label{(50)}
\frac{m}{\hbar} j_+(y) = -\textrm{Re}\left[i \beta^*(y) \left(\frac{d\beta}{dy}\right)\right],
\end{equation}
and the result is a sum of slowly varying terms representing the
measurable depolarization. There are also fast oscillating terms
with phase $ \pm\Phi(y) = \pm[\Phi_+(y)-\Phi_-(y)]$ which would be averaged to zero
by a detector of spin-flipped UCNs
except within a narrow range of order $2\pi/(k_{-m}-k_{+m})$, i.e., of a few wavelengths above the mirror.
With the same proviso,
mixed terms $\sim \xi r^*$ (or $ \xi^* r)$
can also be dropped since the phase factors $\xi$ and $r$
depend sensitively on the exact position of the mirror
and the exact distance between upper and lower turning points.
In practice these quantities are blurred by geometrical imperfections
as well as the finite spread in UCN energy. Thus $\xi$ and $r$
can be considered statistically independent of one another. As a result we obtain for the measurable average current
\begin{align}\label{(51)}
&\Big\langle \frac{m}{\hbar} j_+(y)\Big\rangle  = \frac{k_+}{k_-} |P(y)|^2 + 2 |P(y_m)|^2 \nonumber \\
&= \frac{(k_+\theta')^2+(Kk_x\sin\theta)^2}{(k_-^2-k_+^2)^2}\\
& + 2\left(\frac{k_{-m}}{k_{+m}}\right) \frac{(k_{+m} \theta_m')^2 +
(Kk_x \sin\theta_m)^2}{(k_{-m}^2 - k_{+m}^2)^2} . \nonumber
\end{align}

To measure the depolarization per one complete bounce on the mirror
from upper turning point down and back up to the upper turning point,
we insert $y = y_u$ for the detector position and obtain
 \begin{align}\label{(52)}
&\Big\langle\frac{m}{\hbar} j_+(y_u)\Big\rangle = \left(\frac{Kk_x}{k_{-u}^2}\sin\theta_u\right)^2 \nonumber \\
&+ 2\left(\frac{k_{-m}}{k_{+m}}\right) \frac{(k_{+m} \theta_{m}')^2 + (Kk_x \sin\theta_m)^2}{(k_{-m}^2 - k_{+m}^2)^2}. 
\end{align}
This corresponds to expression (\ref{(36)}) for one bounce in the magnetic
field. In (\ref{(52)}) we have used $k_{+u} = k_+(y_u) = 0$ and we note that in
typical cases the first term in (\ref{(52)}) is negligible.

We have assumed that the UCNs are incident on the mirror from above, i.e. are confined to the space $y_{m}<y<y_{u}$. If, instead, they impinge from below at $y=y_{m}$ (now the lower mirror surface) and are confined to the space $y_{l}<y<y_{m}$, the expressions (\ref{(51)}) and (\ref{(52)}) (now for $j_{-}$, not $j_{+}$) remain the same except that in (\ref{(52)}) the index $u$ is replaced by $l$.

We will discuss these results in greater detail in sections \ref{sec:IV.C} and \ref{sec:VI}.

 \section{\label{sec:IV} SEMI-CLASSICAL APPROACH  }
\subsection{\label{sec:IV.A}    Basic equations}	
	
   The semi-classical Schr\"odinger equation describes the particle in time $t$,
rather than in space coordinates. The particle is assumed
to follow a known classical path, so the field variable $\bm{B}$ is considered a
 known function of $t$. We again choose the quantization axis along
the position-dependent direction of $\bm{B}$, for which the mutually
orthogonal basis vectors $\chi^+$ and $\chi^-$ are given in (\ref{(6)}) and the
wave function for the (+) and ($-$) state is
\begin{equation}\label{(53)}
   \chi(t) = \alpha(t)\chi^+(t) + \beta(t)\chi^-(t).
\end{equation}
But these quantities are now considered to be functions of $t$, rather than of
space variables. As before, the spin flip probability $|\beta(t)|^2$ is considered
small compared to the probability $|\alpha(t)|^2 \approx $ 1 to find the neutron in the
storable (+) spin state.

   In this approximation the Schr\"odinger equation reads \cite{14}
\begin{align}\label{(54)}
& i\hbar\fd{d}{dt}(\alpha\chi^+ + \beta\chi^-) \nonumber \\
& =\mathcal{H}_m(\alpha\chi^+ + \beta\chi^-) = |\mu_\textrm{n}| B (\alpha\chi^+ - \beta\chi^-)
\end{align}
where the Hamiltonian $\mathcal{H}_m$ for spin interaction was given in (\ref{(3)}). We
denote time derivatives by a dot and, using the notation and
relations of Sec.~\ref{sec:III.A} and,
from $\phi=-Kx$, $\dot\phi=v_x \frac{d\phi}{dx} = -v_xK$, obtain
\begin{align}\label{(55)}
&\dot{\chi}^- =\left( \begin{array}{c}
\left(isKv_x+\frac{1}{2}c\dot\theta\right)e_-  \\
\frac{1}{2}s\dot\theta \\
 \end{array} \right)
 \nonumber \\
&=\frac{1}{2}(\dot\theta+iKv_x\sin\theta)e_-\chi^++iKs^2v_x\chi^-
\end{align}
and
\begin{align}\label{(56)}
&\dot{\chi}^+ =\left( \begin{array}{c}
-\frac{1}{2}s\dot\theta  \\
\left(\frac{1}{2}c\dot\theta-isKv_x\right)e_+\\
 \end{array} \right) \nonumber \\
&=-\frac{1}{2}(\dot\theta-iKv_x\sin\theta)e_+\chi^--iKs^2v_x\chi^+ .
\end{align}

Inserting into (54) and keeping only the dominant contributions we get for the terms with $\chi^+$
\begin{equation}\label{(57)}
\dot\alpha+\fd{i\omega_{L}}{2}\alpha=0
\end{equation}
and for those with $\chi^-$
\begin{equation}\label{(58)}
\dot\beta-\fd{i\omega_{L}}{2}\beta=\fd{\alpha}{2}(\dot\theta-iKv_x\sin\theta)e_+ ,
\end{equation}
where $\omega_{L}=2|\mu_{\textrm{n}}|B/\hbar$ is the Larmor frequency. In practice, $\omega_{L} \gg Kv_x$. Equations (\ref{(57)}-\ref{(58)}) correspond to equations (14) and (15) of \cite{14}, but (\ref{(58)})
contains the new phase factor $e_+ =\exp(i\phi) = \exp(-iKx)$ (as the quantum
equivalent, Eq.~(\ref{(12)}), does)
and the term dependent on $v_x$.

\subsection{\label{sec:IV.B}    Depolarization in the magnetic field}	

   Equation (\ref{(57)}) is solved by
\begin{equation}\label{(59)}
	\alpha(t) = \exp\left(-\frac{i\Theta}{2}\right)
\end{equation}
where $\Theta = \int_{t_s}^t \omega_{L}(t') dt'$ is twice the phase
angle accumulated since the  start time $t_s$.  As for the quantum case,
we assume that the motion starts at the upper or lower turning point with
velocity $\bm{v}_s = (v_x, v_{+s}, v_z) = (v_x, 0, v_z)$, where $v_x =$ const. and the
constant $z$-component $v_z$ does not induce depolarization for our field
model. $v_{+s} = v_+(t_s) = 0$ implies $\dot\theta = 0$ at the start.

   As in \cite{14}, Eq.~(\ref{(58)}) is solved using the ansatz  $\beta(t) = G(t)\exp(i\Theta/2)$, where $G(t)$,
the new measure of depolarization amplitude \cite{14}, satisfies the relation
\begin{equation}\label{(60)}
\dot G = \frac{1}{2}(\dot\theta - iKv_x \sin\theta) \exp(-i\Theta) \exp(-iKx).
\end{equation}
Since $x = v_xt$ the factor $\exp(-iKx) = \exp(-iKv_xt)$ is time dependent and
therefore $i\hbar \fd{d}{dt}    \exp(-iKv_xt)$ contributes to the energy.
However, as noted before, $Kv_x$ is much smaller than $\omega_L$ and can be neglected in the same way
small terms are neglected in the WKB approximation.

   Eq.~(\ref{(60)}) is readily integrated by parts:
\begin{align}\label{(61)}
&G(t) =\int_{t_s}^t \dot G(t') dt'\nonumber \\
& = \frac{i}{2\omega_L} (\dot \theta - iKv_x \sin\theta) \exp(-i\Theta)\exp(-iKx).
\end{align}
As for the quantum analog (\ref{(28)}) of (\ref{(68)}), the right-hand side of Eq.~(\ref{(61)})
does not include a term for the lower integration limit $t_s$. As we will see, this
leads to results matching those of the quantum approach. We again can argue,
as in the discussion of Eq.~(\ref{(28)}), that in view of the Airy-function character of the
actual wave solution around the turning region, the initial time $t_s$ is a blurred
quantity and the contribution from the lower limit of integration averages to zero.
But a rigorous justification may be impossible within this semi-classical mix of ingredients as incongruent as classical and quantum mechanics are.

   We identify $|G(t)|^2$ with the probability $p$ of finding the neutron in the
spin-flipped state \cite{14}
\begin{equation}\label{(62)}
p = \frac{\dot \theta^2+K^2v_x^2 \sin^2\theta}{4\omega_L^2}=\frac{k_+^2 {\theta'}^2 +K^2k_x^2\sin^2\theta}{(k_-^2 - k_+^2)^2} .
\end{equation}
In the last step of Eq.~(\ref{(62)}) we have used $\omega_L = \frac{\hbar}{2m}(k_-^2 - k_+^2)$,
and  $\dot\theta^2=v_+^2\left(\frac{d\theta}{dy}\right)^2 =\left(\frac{\hbar}{m}\right)^2 k_+^2{\theta'}^2$, since in our field model $\theta$ depends only on $y$.

   The result (\ref{(62)}) agrees with Eq.~(\ref{(34)}) for the function $(m/\hbar) j_- $
 which had
also been identified as the depolarization probability. Therefore we have full
agreement also for the mean depolarization per bounce (\ref{(37)}) and
for the depolarization rate given in Eq.~(\ref{(39)}).

   In the next section we will see that the semi-classical and quantum approaches
do not always produce exactly the same results, although a strong correlation
between the two will be found also in this case.

\subsection{\label{sec:IV.C}    	Reflection from a non-magnetic mirror in a magnetic field}

  The semi-classical analysis of the mirror problem follows the same path as for
the quantum case analyzed in Sec.~\ref{sec:III.C}. The crucial step again is matching the superposition
of particular plus homogeneous solution for the ascending reflected beam,
\begin{equation}\label{(63)}	
			\beta(t) = \beta_{p}(t) + \beta_{h}(t),
\end{equation}
to the wave $\beta_m$ induced by the incident beam at the mirror surface at $t=t_m$.

In Eq.~(\ref{(63)}), $\beta_{h}=C\exp\left(\frac{-i\Theta}{2}\right)$ is the general solution of the homogeneous equation $\dot{\beta}_{h}+\frac{i\omega_{L}}{2}\beta_{h}=0$, which corresponds to the inhomogeneous equation (\ref{(58)}) with the direction of time reversed as described following Eq.~(\ref{(44)}). Using the same sequence of terms as in (\ref{(63)}) and the notation of Sec.~\ref{sec:III.C}, the matching condition $\beta(t_{m})=\beta_{p}(t_{m})+\beta_{h}(t_{m})$ reads
\begin{align}\label{(64)}
&\frac{-i\zeta_1}{2\omega_{Lm}} (\dot\theta_m + iKv_x \sin \theta_m)\exp\left(+\frac{i\Theta_m}{2}\right)
\nonumber \\
&= \frac{-i\zeta_2}{2\omega_{Lm}}(\dot\theta_m+iKv_x\sin\theta_m)\exp\left(+\frac{i\Theta_m}{2}\right)\\
& + C\exp\left(-\frac{i\Theta_m}{2}\right), \nonumber
\end{align}
where we have divided out the common factor $\exp(iKv_xt_m)$. $\omega_{Lm}$
is the Larmor frequency at the mirror position and $\zeta_1$, $\zeta_2$ are statistically independent unitary phase factors similar to their
quantum analogs $\xi$, $r$ in (\ref{(44)}) and (\ref{(46)}). Solving for the constant $C$ of the
homogeneous term and inserting into (\ref{(63)}) gives
\begin{align}\label{(65)}
&\beta(t) =-\exp(iKv_xt)\Bigg[\frac{i}{2\omega_L}(\dot\theta+iKv_x\sin\theta)\exp\left(+\frac{i\Theta}{2}\right)+
 \\
&\frac{i(\zeta_1-\zeta_2)}{2\omega_{Lm}}(\dot\theta_m + iKv_x\sin\theta_m)\exp(+i\Theta_m) \exp\left(-\frac{i\Theta}{2}\right)\Bigg].\nonumber
\end{align}
   Finally, taking the square magnitude of (\ref{(65)}) and performing the same statistical
averaging as for (\ref{(50)}), which includes setting $\Big\langle |\zeta_{1}|^{2} \Big\rangle=\Big\langle |\zeta_{2}|^{2}\Big\rangle=1$ and $\Big\langle \zeta_{1}\zeta_{2}^{*} \Big\rangle=0$, we find for the probability of depolarization
\begin{align}\label{(66)}
&p(t) = \Big\langle |\beta(t)|^2 \Big\rangle = \frac{1}{4\omega_L^2} \left[\dot\theta^2 + (Kv_x\sin\theta)^2\right]\nonumber \\
&+ \frac{1}{2\omega_{Lm}^2} \left[\dot\theta_m^2 + (Kv_x\sin\theta_m)^2\right].
\end{align}
To facilitate comparison with the quantum result (\ref{(51)}) we convert from time to space dependent variables $\left(\dot\theta \rightarrow\theta' = \frac{d\theta}{dy}\right)$, as for
Eq.~(\ref{(62)}), and obtain
\begin{align}\label{(67)}
&     p(t) = \frac{(k_+\theta')^2+(Kk_x\sin\theta)^2}{(k_-^2 - k_+^2)^2} \nonumber \\
&+\frac{2[(k_{+m} \theta_{m}')^2 + (Kk_x \sin\theta_m)^2]}{(k_{-m}^2 - k_{+m}^2)^2}.
\end{align}
 This is the same expression as (\ref{(51)}) except that the factor $k_{-m}/k_{+m}$ for
the second term on the right-hand side is missing. Except for very
low energy UCNs hovering in the magnetic field,
 and for mirror position at a turning point, this factor is close to 1.

   For a complete bounce on the mirror, starting from,
and ending at, the upper turning point level $y_u$ we obtain
\begin{equation}\label{(68)}
p = \frac{(Kk_x\sin\theta_u)^2}{k_{-u}^4}+\frac{2[(k_{+m}\theta'_m)^2+(Kk_x\sin\theta_m)^2]}{(k_{-m}^2-k_{+m}^2)^2} .
\end{equation}

\begin{figure}[tb]
  \begin{center}
 \includegraphics[width=77mm]{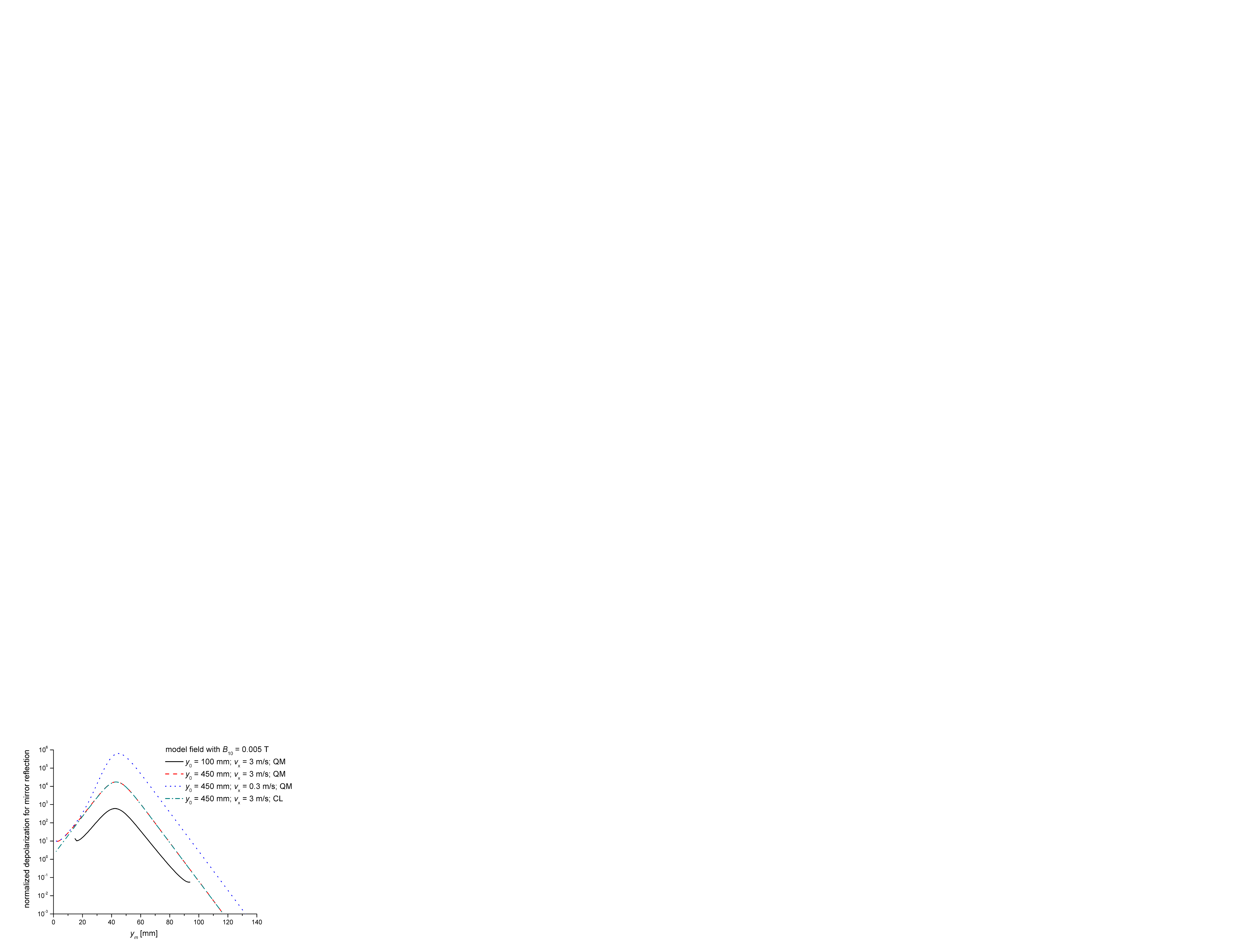}
\end{center}
\caption{(color online) Normalized depolarization per bounce as a function
of mirror position $y_m$.
We use Eq.~(\ref{(52)}) for the quantum treatment (QM) and Eq.~(\ref{(68)}) for
the semi-classical approach. The plotted values are normalized by
dividing (\ref{(68)}) by (\ref{(36)}), the depolarization due to the field alone, and by the ratio of duration of one bounce with and without the mirror. The data show a strong enhancement of depolarization due to the
mirror. The enhancement factor depends on incident UCN energy
(which increases with larger fall height $y_0$) and on the
gradient $\theta_m'$ of field angle $\theta$ at the mirror position.
$\theta_m'$  is
 largest in the region where the curves have their peak
value which is of order $10^4$ for $v_x$ = 3 m/s. For the examples shown, the turning point levels are: $y_{u}=96.86$ mm, $y_{l}=14.31$ mm for $y_{0}=100$ mm and $y_{u}=445.82$ mm, $y_{l}=0.58$ mm for $y_{0}=450$ mm.}

  \label{fig:five}
\end{figure}

   Fig.~\ref{fig:five} shows the depolarization per bounce on the mirror in our model field
as a function of mirror position $y_m$. For the quantum treatment (QM) this
probability is given by Eq.~(\ref{(52)}) and for the semi-classical approach (CL)
by Eq.~(\ref{(68)}). To separate the role of the non-magnetic mirror in the field from the
depolarization due to the field alone we have divided (\ref{(68)}) by (\ref{(36)}), the
depolarization due to the field alone, and by the ratio of duration of one bounce with and without the mirror. Fig.~\ref{fig:five} shows a strong enhancement due to
the mirror. The enhancement factor depends on incident UCN energy (which
increases with larger fall height $y_0$) and on the gradient $\theta_m'$
of field angle at the mirror position. $\theta_m'$  is
largest in the
region where the curves  have their peak.  The enhancement  reaches four orders
of magnitude for $y_0 = 0.45$ m and $v_x = 3$ m/s, and six decades for $v_x = 0.3$ m/s.
In the latter comparison, the role of lateral velocity component $v_x$ appears so
large because the depolarization in the field alone vanishes for $v_x = 0$ in our
approximation (\ref{(36)}) (and is negligibly small in higher-order approximations \cite{14}),
while the mirror depolarization remains finite for $v_x = 0$.

    The quantum and semi-classical results, compared for identical parameters by
the dashed and dash-dotted curves in Fig.~\ref{fig:five}, are quite similar. For the given parameters they differ by a few percent at most if we exclude the range within $\sim$ 5 mm from a turning point.

As for the quantum approach, the results (\ref{(67)}) and (\ref{(68)}) remain the same (except for the index $u$ in (\ref{(68)}) changing to $l$) if incidence on the mirror from above is replaced by incidence from below, with start from the lower turning point $y_{l}$ rather than from $y_{u}$.

   In the next section we will show  that for low UCN energies these analytical
results can also be obtained by numerical integration of the basic differential
equation. Numerical integration does not rely on the WKB approximation.
However, the lengthy integration over many oscillations of the wave function
is plagued with the rounding errors and the error due
to the finite step size. Reasonable agreement of the two methods would be
an indication that the results are reliable.

\section{\label{sec:V}     NUMERICAL INTEGRATION}

\subsection{\label{sec:V.A}	Magnetic confinement}

   The equation of motion of the wave function $\beta(y)$ for spin flip is given by the
second-order inhomogeneous differential equation (\ref{(23)}),

\begin{equation}
\beta''(y) + k_-^2(y) \beta(y) = [\pm ik_+\theta'(y)+ Kk_x \sin\theta(y)] \alpha(y),
\nonumber
\end{equation}
 with (\ref{(16)})
\begin{equation}
    \alpha(y) = k_{+}^{-1/2}(y)\exp\Big(\pm i\Phi_+(y)\Big)
\nonumber
\end{equation}
and the + ($-$) sign refers to motion upward (downward). Using the fourth-order Runge
 Kutta process, Eq.~(\ref{(23)}) may be integrated numerically,
 starting from an initial point $y_i$
slightly above (below) the turning point $y_s$. Key is the suitable choice of initial
values $\beta(y_i)$ and $\beta'(y_i)$.

  We choose the WKB solution which is given by (\ref{(32)}) for downward motion. Its extension
 to include also the upward path reads
\begin{equation}\label{(69)}
			\beta(y) = k_{-}^{-1/2}(y)P_{\pm}(y) \exp\Big(\pm i \Phi_+ (y)\Big)
\end{equation}
with derivative
\begin{equation}\label{(70)}
  \beta'(y) = \pm ik_+(y) k_{-}^{-1/2}(y)P_{\pm}(y) \exp\Big(\pm i\Phi_+(y)\Big),
\end{equation}
where $P_-(y) = P(y)$ and $P_+(y) = P^*(y)$. $P(y)$ has been defined in (\ref{(30)}). As discussed
 following (\ref{(19)}), near a turning point $y_s$ Eq.~(\ref{(69)}) is based on the asymptotic form of the Airy function wave
solution $\alpha(y) = C_1 \textrm{Ai}(-a_s|y - y_s|)$, where the coefficient
$C_1 = 2(\pm i\pi /a_s)^{1/2}$ is adjusted to match (\ref{(16)}) asymptotically. The constants $g_{+s}$ and $a_{s}$ were defined following Eq.~(\ref{(19)}). Since $g_+$ varies slowly,
the constant value $g_{+s}$ is a good approximation over hundreds of oscillations
of the Airy function, starting from $y = y_s$.

\begin{figure}[tb]
  \begin{center}
 \includegraphics[width=77mm]{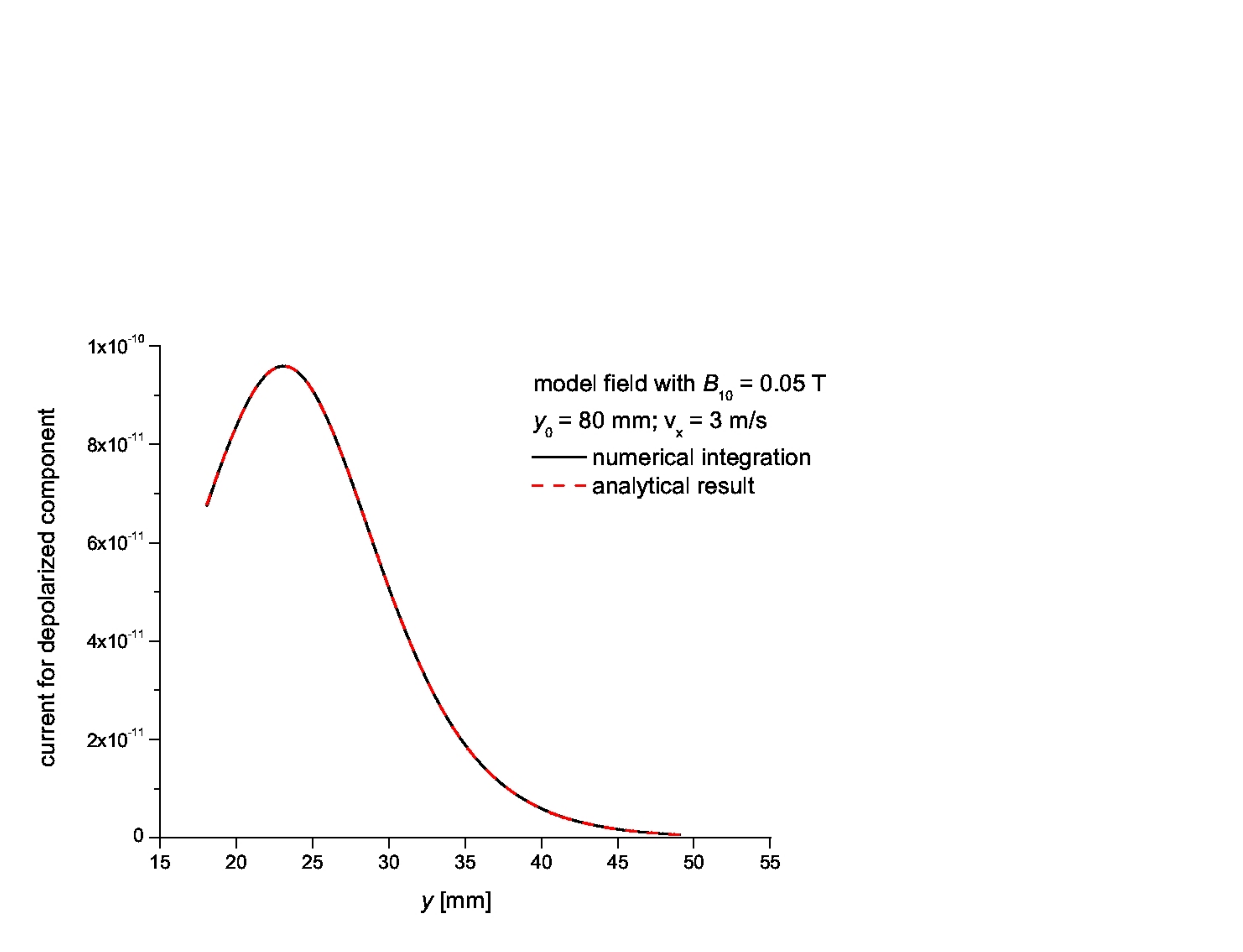}
\end{center}
\caption{(color online) Depolarization current for downward motion in the magnetic
field for parameters $y_0$ = 8 cm, $B_{10}$ = 0.05 T and $v_x$ = 3 m/s.
Direct numerical integration, represented by the solid curve, coincides with the analytic result shown by the dashed curve,
with a maximum deviation of 1\% over the entire range
from upper to lower turning point ($y_{u}=49.56$ mm, $y_{l}=18.08$ mm).}
  \label{fig:six}
\end{figure}

   Using initial values (\ref{(69)}) and (\ref{(70)}), Eq.~(\ref{(23)}) may be integrated numerically
 up to a point slightly before the next turning point, $y_{s,\textrm{next}}$, is reached. The
method fails at $y_{s,\textrm{next}}$ itself due to the divergence of $k_+^{-1/2}$. We calculate
the current $(m/\hbar) j(y) =\pm{\textrm{Re}}[\beta^*(y) \beta'(y)/i]$ at every point
along the integration
 path and Fig.~\ref{fig:six} shows the result of downward integration for parameters
$ y_0 = 8$ cm, $B_{10} = 0.05$ T, $v_x = 3$ m/s and starting point
at $|y_i - y_s| = 10/a_s$. We have tested that the solution is stable
 in a wide range of initial position from $a_s|y_i - y_s| \sim$2 to 20.

   Fig.~\ref{fig:six} shows that the numerical result coincides with the analytical
solution (\ref{(69)}), with maximum deviations of $\sim1\%$ over the entire range
 including the far endpoint $y_l$. This consistency at the lower turning point
 is vital since the current leaving there from the storage space is identified
with the depolarization probability for the move from top to bottom. To be
specific, this depolarization probability $(m/\hbar) j_-(y_l)$ is obtained by extrapolating
the numerical solution over the short distance of order $a_{s,\textrm{next}}^{-1}$ to
the next endpoint $y_{s,\textrm{next}}$ with the result  $(m/\hbar) j_-(y_l) =
6.80\times 10^{-11}$ for the case shown in
 Fig.~\ref{fig:six}. The extrapolation at $y_l = 18.08$ mm is straightforward since the current
shows a smooth behavior in the entire integration range. The same level of
 agreement within $\sim 1\%$ is obtained for the upward motion which is
represented by the same curve of Fig.~\ref{fig:six}, but the extrapolation for
the loss current is now made at the upper end $y_u = 49.08$ mm
(and this contribution is much smaller in the case shown, but it is
comparable for smaller drop heights). We found agreement
within 1\% between the various methods also for other parameters
 $y_0, B_{10} \textrm{, and  } v_x$ within the range of interest.

   There is one exception. For the strictly vertical mode of motion
 analyzed in Ref.~\cite{14}, where $v_x = 0$, the analytical solution vanishes
 and the numerical solution for the intercept is at least four orders of
magnitude smaller than for $v_x = 3$ m/s. In this case the numerical precision is
insufficient to determine a reliable value of depolarization probability. However, this
 is inconsequential since extremely small values of $v_x$ make a negligible contribution
to the mean depolarization rate in a broad UCN spectrum.

   We add one remark. If the explicit analytical solutions (\ref{(69)}-\ref{(70)}) for $\beta$ and $\beta'$
were not known, we could use, for the numerical integration, initial values
 derived solely from the properties of the Airy function solution near
the turning points. In these regions the second derivative $\beta''$ in Eq.~(\ref{(23)}) is
negligible compared with $k_-^2\beta$ since
 $k_-^2 \gg a_s^3|y-y_s|=k_+^2$.
Therefore,
 $\beta(y) \approx -[\pm ik_+\theta'(y) - Kk_x \sin\theta(y)] \alpha(y)/k_-^2(y)$
 is a good approximation.
 It differs from (\ref{(69)}) only by the multiplier $[1-k_+^2(y)/k_-^2(y)]$ which
is close to 1 at either turning point since $k_+$ vanishes there. It turns
out that these modified initial values give the same results
for $\beta(y)$ at both endpoints, and therefore the same depolarization
probability, as the more exact method. However, the function $\beta(y)$
will be somewhat different in the range between the endpoints, where $\beta''$ cannot be neglected.

   We may summarize the results on depolarization in magnetic
storage as follows: The fact that the three methods used agree
(quantum and semi-classical analysis as well as direct numerical
integration) appears to be a good indication that the
 approximations made were justified.

\subsection{\label{sec:V.B}	Mirror reflection}

   For UCN reflection from a non-magnetic mirror in a magnetic
 field the analytical results given in (\ref{(52)}) and (\ref{(68)}) are similar but not
 identical. We used numerical integration for the reflected beam, starting
from initial conditions (\ref{(46)}) at the mirror surface and note that in this
case the extrapolation to the next turning point is less straightforward
since the curve for current shows, superimposed on a smooth variation, fast oscillations due to
 beating between two wave components: one propagating with wave
 number $k_+$ and the other with $k_-$. This is expected for the superposition
 of these two waves in Eq.~(\ref{(48)}). The numerical results are generally
consistent with the analytical expressions but not precise enough to decide
 whether the semi-classical approximation (\ref{(68)}) or the quantum
 approximation (\ref{(52)}) is more reliable.

\section{\label{sec:VI}    DISCUSSION}

Depolarization in high precision neutron lifetime experiments using magnetic confinement must either be negligible or else very small and quantitatively understood. Using two analytical methods based on Ref.~\cite{14} and direct numerical
 integration we have analyzed the depolarization per bounce and the depolarization
 rate (per s) for UCNs stored in a model magnetic field configuration. Our magnetic field model is similar to
the system envisaged for the ``bathtub'' project \cite{14} which uses a
Halbach array of permanent magnets. Our model is simplified to a
configuration with translational symmetry in both horizontal
directions ($z$ and $x$) and an ideal Halbach field whose magnitude depends
only on the vertical distance $y$ from the horizontal magnet surface. The
parameters for the Halbach field are the same as those proposed
 in Ref.~\cite{14}. Our analysis  shows that
depolarization is mainly caused by the rotating Halbach
field the UCNs see as they move through the field with finite
lateral velocity component $v_x$, not by the small field ripple due to
imperfections of the Halbach system. Therefore we do not expect
 the simplification of the model to affect the depolarization in a
significant way.

   However, the role of the additional horizontal stabilization
field $\bm{B}_1$ perpendicular to the Halbach field $\bm{B}_H$ is critical. It has
to be strong enough to suppress depolarization to an
acceptable level. The main purpose of this work was to
determine tolerance limits for its magnitude $B_1$.

   Our analysis extends that of Ref.~\cite{14} by including arbitrary
 UCN orbits in 3D space whereas the analysis in \cite{14} was restricted
 to purely vertical motion. As a main result of the extension we
find that the lateral $x$-component of motion in the plane of the
Halbach field makes the dominant contribution to depolarization
while the depolarization due to the vertical motion is insignificant.
As a result, some previous estimates of depolarization probability
may have been overoptimistic. For the parameters
of \cite{14} (0.05-0.1 T for $B_{10}$) we estimate on the basis of Fig.~\ref{fig:three}
that even a measurement of the neutron lifetime with precision
 $10^{-5}$ should be possible (disregarding other potential limitations) but the safety margin may be smaller
 than previously expected.

   Systems with smaller stabilization field, as possibly that
of Ref.~\cite{11}, where $B_1$ has not been specified, may require
 a separate analysis since they use a cylindrical rather than
planar field distribution. However, the main result of the
present work is independent of geometrical details: The
depolarization loss is determined, not by the largest rotation
frequency of $\bm B$ as seen by the UCNs as they traverse the field,
but by the conditions at the turning points where the spin-flipped neutrons can leave the storage system.

   This is an important point which may appear to contradict the
common view, but is also implied by the work of Ref.~\cite{14}. Our
interpretation of the present results is as follows: The measurable
 depolarization is not directly caused by critical spots within
the storage volume, where the $\bm B$-field has a small magnitude $B$
and rotates fast in the moving reference frame. All that matters
for depolarization are the field conditions at the turning points
where the storable UCNs are reflected back into the storage space
while the spin-flipped fraction leaves the trap. The loss current
 in Eq.~(\ref{(37)}) is determined by the value of $K^{2}k_{x}^{2}\sin^2\theta/k_-^4$ at such a
surface. This factor is large for fast field variation ($\omega\sim v_{x}/L \sim Kk_{x}$) seen by the neutron moving in a horizontal direction (parallel to the turning surface), as well as for small $B$ since $k_-^2$ from (\ref{(31)}) is
 directly proportional to $B$ at a turning point where $k_+ = 0$. It also
 increases quadratically with sin$\theta = B_{H}/B$, i.e. with the magnitude $B_{H}$ of the Halbach field at a turning surface. If a broad spectrum of UCNs is stored the critical points
 will be of importance in the sense that some, usually
UCNs with very low energy of vertical motion,
 may have a turning surface near such areas
and therefore make a large contribution. In fact, the low-energy
 UCNs make the largest contribution to depolarization seen in the
peaks in Fig.~\ref{fig:three}, and they are the reason why a larger stabilization
 field $B_1$ suppresses the net depolarization very effectively.

   Fig.~\ref{fig:three} shows $\tau_{\textrm{dep}}^{-1}/\tau_{\textrm{n}}^{-1}$,
the depolarization rate divided by
the $\beta$-decay rate, as a function of normalized vertical
velocity $v_+^{(\textrm{n})}$ in the neutral plane, where gravity and the
 magnetic force are balanced. Both curves, for $B_{10}$ = 0.05 T
and for $B_{10}$ = 0.005 T, are peaked at small values of $v_{+}^{\textrm{(n})}$. Thus the UCNs most in danger of
suffering depolarization are confined to a narrow space
of $\sim$10 cm about the neutral plane, moving laterally as
they float in the field (for $v_+^{\textrm{(n)}} = 0)$ or oscillate up and down
about the neutral plane with small amplitude.

The magnetic field configurations and neutron orbits in actual or projected 3D magneto-gravitational UCN confinement systems \cite{11,12,13,14,15} are more complex than in our field model. However, for some of these concepts the 1D approximation of our model appears to be justified. For the ``bathtub" system \cite{14}, the neutral plane of our model, about which low-energy UCNs oscillate, corresponds to a strongly anisotropic oscillator potential in the vicinity of the minimum of potential $gy+|\mu_{n}|\frac{B}{m}$ for the low-field seeking spin state. For $B_{10}=0.005$ T the minimum is located about 1.8 cm up from the lowest point of the double-curved surface of permanent magnets of Ref.~\cite{14}. For this anisotropic oscillator, the frequency for oscillations in the vertical $y$-direction, $\omega_{0y}=\left(\frac{dg_{+}}{dy}\right)^{1/2}=57$ s$^{-1}$, is 22 times larger than for the $z$-direction, $\omega_{0z}=\left(\frac{g}{R_{z}}\right)^{1/2}= 2.6$ s$^{-1}$ where the radius of curvature is $R_{z}=\rho=1.5$ m. For the $x$-direction, with its asymmetry, there are two curvatures (0.5 and 1.0 m) and, therefore, two ratios replacing 22: 13 on one side ($x<0$) and 18 for $x>0$. Since all these factors are large, the neutrons move almost freely, on a relative scale, in the peripheral $x$- and $z$-directions and, therefore, our 1D model should be a good approximation. As a result, we expect the peripheral velocity in the plane of the Halbach field, which corresponds to $v_{x}$ in the model, to be the main source of depolarization.

In the cylindrical field geometries of references \cite{11,12,13,15}, the field magnitude $B$ varies more slowly in space. Thus, the oscillator is less anisotropic, the degrees of freedom of motion in different coordinate directions are less decoupled and a more complex analysis may be required. To the extent that qualitative features of our model may still apply we expect that the main source of depolarization would be a large peripheral UCN velocity perpendicular to the cylinder axis. In this case, the peripheral velocity corresponds to the component $v_{x}$ of the model.

   Besides depolarization of magnetically confined UCNs
 we also studied depolarization in UCN reflection on a
non-magnetic mirror immersed in a magnetic field. The
 field was our model field into which we conceptually inserted
 an ideal neutron mirror horizontally at a variable height. The
 problem of possible depolarization in mirror reflection is of
 paramount importance in UCN experiments on the asymmetry
 parameter $A$ in neutron decay \cite{20,21} and it is also encountered
 in high-field UCN polarizers. Depolarization is expected since the adiabaticity condition may be violated due to the abrupt change of flight direction at the reflection point, thus $d{\mathbf{B}}/dt$ changes abruptly. This problem has first been studied in Ref.~\cite{17} by
 adapting the Majorana semi-classical approach \cite{18} to the mirror
geometry. In our analysis the three methods used (quantum
approximation, semi-classical and numerical approach) gave identical
 results for depolarization in magnetic UCN storage. For the mirror
 reflection problem the quantum result (\ref{(52)}) and the semi-classical
result (\ref{(68)}) are very similar but not identical, as shown in Fig.~\ref{fig:five}.
The numerical method is not accurate enough in this case to distinguish
between the two. The semiclassical result (\ref{(68)}) lacks the factor $k_{-m}/k_{+m}$ which would cause a
 divergence if the mirror is placed at a turning point height
(since $k_{+}=0 $ in this region).

   Between the turning points the difference is minor and the
common result is as follows: Depolarization in mirror reflection, averaged over field
directions as in our model field, is determined mainly by the second term of Eq.~(\ref{(68)}). It increases with the frequency of field variation ($\sim Kk_{x}$) seen by the neutron moving along the in-plane $x$-direction. It also increases quadratically with the sine of the field
angle $\theta$, which is a measure of Halbach field strength $B_{H}=B\sin\theta$, and with its gradient, $\theta' = \frac{d\theta}{dy}$, at the mirror location. Depolarization strongly decreases with increasing field strength $B$ at the mirror ($\sim B_{H}^{2}/B^{4}$). Fig.~\ref{fig:five} shows that, for $v_x$ = 3 m/s, the
 magnitude of depolarization per one bounce on the mirror is up to $\sim 10^4$ times larger than the
 depolarization per bounce in the magnetic field without the mirror.
 The depolarization on the mirror has
its peak value at the vertical location where the
depolarization probability plotted in Fig.~\ref{fig:two} also has its peak. In fact, comparing expression (\ref{(68)}) for depolarization at the mirror
 with the result (\ref{(34)}) for the field alone (and neglecting the small first term on the right-hand side of (\ref{(68)})) we realize that the mirror
 acts like a polarization analyzer inserted into the particle beam
 moving through the {\bf{B}}-field. This interpretation also holds for a non-horizontal or curved mirror since the second term on the right-hand side of (\ref{(68)}) is independent of the orientation of the reflecting surface element. In this case, the second term should be averaged over the mirror extension.

    In cases where many successive wall reflections take place in a weak, non-uniform magnetic
field, the depolarization may become significant.
Comparing our result (\ref{(66)}) with equations (10-11) of Ref.~\cite{17} we
note that  both results have the square of the Larmor frequency $\omega_{L}$ at the mirror position in the denominator. However, a quantitative
comparison is difficult since the magnetic field variations assumed
 in the two approaches are different. In either analysis, no depolarization is expected for a uniform magnetic field.

Our field model may be too specific to allow a quantitative comparison with the data \cite{25,26} on depolarization in UCN reflection from various mirror materials (like beryllium or samples with diamond-like carbon coating). Expression (\ref{(68)}) does not depend on specific properties of the mirror, as long as it is a good, nearly loss-free UCN reflector. Therefore, (\ref{(68)}) could explain, without having to invoke any anomalies \cite{25}, the remarkable similarity and temperature independence of depolarization probabilities measured for different wall materials. Such independence would be expected if the samples were exposed to the same non-uniform magnetic field.

Our results, which were obtained as straightforward solutions to the spin-dependent Schr\"odinger equation, may also provide an alternative to the discussion of new short-range, spin-dependent forces as a possible pathway to explaining the depolarization data for stored UCNs. For a recent comprehensive review of fundamental physics with neutrons see \cite{27}.

\begin{acknowledgments}

We thank R. Golub and B. G. Yerozolimsky for helpful comments and V. Ezhov, C. Liu and A. Young for having directed our attention to the
question of depolarization in magnetic UCN confinement.

\end{acknowledgments}

\pagebreak

\end{document}